\documentclass[prl,twocolumn,superscriptaddress,noshowpacs]{revtex4}
\usepackage{graphicx}
\usepackage{physics}
\usepackage{bm}
\usepackage{ulem}
\usepackage{color}

\begin{document}

\title{Domain-wall topology induced by spontaneous symmetry breaking in polariton graphene}

\author{D.~D.~Solnyshkov}
\affiliation{Institut Pascal, PHOTON-N2, Universit\'e Clermont Auvergne, CNRS, Clermont INP, F-63000 Clermont-Ferrand, France.}
\affiliation{Institut Universitaire de France (IUF), F-75231 Paris, France}
\author{C.~Leblanc}
\affiliation{Institut Pascal, PHOTON-N2, Universit\'e Clermont Auvergne, CNRS, Clermont INP, F-63000 Clermont-Ferrand, France.}
\author{I.~Septembre}
\affiliation{Institut Pascal, PHOTON-N2, Universit\'e Clermont Auvergne, CNRS, Clermont INP, F-63000 Clermont-Ferrand, France.}
\author{G.~Malpuech}
\affiliation{Institut Pascal, PHOTON-N2, Universit\'e Clermont Auvergne, CNRS, Clermont INP, F-63000 Clermont-Ferrand, France.}

\begin{abstract}
We present a numerical study of exciton-polariton (polariton) condensation in a staggered polariton graphene showing a gapped s-band at the $K$ and $K'$ valleys. The condensation occurs at $K$ or $K'$, at the kinetically-favorable negative mass extrema of the valence band. Considering attractive polariton-polariton interaction allows to generate a spatially extended condensate.
Spontaneous symmetry breaking occurring during the condensate build-up leads to the formation of valley-polarized domains following the Kibble-Zurek scenario. The selection of a single valley breaks time-reversal symmetry and the walls separating domains exhibit a topologically-protected chiral current. This current therefore emerges as a result of the interplay between the non-trivial valley topology and the condensation-induced symmetry breaking.
\end{abstract}


\maketitle

Topological physics is now a well-developed field taught in the Universities. It has changed our understanding of physical systems and brought new approaches to their description. It is now clear that topologically-nontrivial crystals actually represent a significant fraction \cite{Vergniory2019}, and not a rare delicate case in solid state physics. Besides the fundamental change of the paradigm, topological physics also brings new applications ranging from topologically-protected qubits \cite{DasSarma2005,Gladchenko2009} to topological lasers \cite{Bahari2017,Bandres2018} and optical isolators \cite{Solnyshkov2018,Karki2019}.

One peculiar type of lattice hosting topological phases are staggered honeycomb lattices,  implemented naturally in boron nitrides \cite{novoselov2005two}, transitional metal dichalcogenides \cite{novoselov20162d}, but also in artificial optical lattices  \cite{Noh2018}. In general, honeycomb lattices show two Dirac valleys called $K$ and $K'$ at the corner of their hexagonal Brillouin zone. The staggering opens a gap. Each valley can be approximately described by a 2D massive Dirac Hamiltonian. The corresponding states are characterized by non-zero angular momentum and Berry curvature, which can further be linked with a valley Chern number $\pm1/2$ in both valleys respectively. The total Chern number including both valleys is zero and a staggered honeycomb lattice is topologically trivial as a whole, but it can still host a quantum pseudo-spin Hall phase called quantum valley Hall effect \cite{Niu2007,Yao2009}. Indeed, making a zigzag interface between two lattices with opposite staggering creates two interface states whose propagation direction is linked with the valley and related to the difference between valley Chern numbers \cite{Ma2015,ju2015topological,Noh2018}.

On the other hand, the band structure of a solid, topological or not, is obtained in a single-particle approximation: the interactions between the particles and their nature are neglected. However, the description of a quantum fluid in a lattice is a many-body problem. For fermions, one is generally interested in the behavior of the Fermi surface, which gives rise to such effects as Fermi arcs in Weyl semi-metals \cite{xu2015discovery}. In the simplest approximation, the Fermi surface is obtained by filling the single-particle states with electrons, neglecting their interactions, justified by the Pauli exclusion principle \cite{AshcroftMermin}. For bosons, the formation of a Bose-Einstein condensate in a lattice has also been shown to lead to spectacular topological effects \cite{wu2016realization}. Here, the basic case is the formation of a condensate in a particular single-particle state, with the possibility to calculate the topology of the condensate's weak excitations (bogolons) \cite{Engelhardt2015,Furukawa2015,aidelsburger2015measuring}, where the interactions can lead to topological transitions \cite{HadadNatElec2018,maczewsky2020nonlinearity}. 
Other situations include the formation of purely non-linear solutions (e.g. solitons) \cite{Lumer2013,Solnyshkov2017,Smirnova2020}, often bifurcating from linear topological states \cite{kirsch2021nonlinear,mukherjee2020observation,Guo2020}. In these cases for bogolons and solitons, the topology of the lattice is inherited by the non-linear states \cite{Furukawa2015,kirsch2021nonlinear}. The influence of the lattice symmetry breaking by the laser phase on the non-linear topological gap solitons was shown quite recently \cite{pernet2021topological}. It was also shown that the quantum fluid can  bring in its own topology, which can reinforce the protection provided by the lattice \cite{Bleu2018nc}. 

The topology of the bosonic quantum fluid is due to the complex-numbered nature of the wave function, which can thus be decomposed into an amplitude and a phase. The phase winding number is a topological invariant \cite{Thouless1998} protecting the quantum vortices. Their topological protection plays an important role in the Kibble-Zurek mechanism \cite{Kibble1976,Zurek1985,Zurek1996} (KZM), which consists in the formation of domains of the order parameter during second-order phase transitions, such as the Bose-Einstein condensation. These domains then decay into topologically-protected defects, whose density can be measured. The walls, separating these domains, can also behave as topological defects \cite{yao2022domain}, lasting as long as the domains they surround.

In this work we study the dynamics of polariton condensation in a uniform staggered honeycomb lattice (without quantum valley Hall interface). We show that condensation is kinetically favored in the negative mass $K$ and $K'$ states being at the top of the valence band. We consider attractive polariton-polariton interactions allowing the formation of a spatially homogeneous condensate density. During the condensate formation, a spontaneous symmetry breaking by phase fluctuations forms valley-polarized spatial domains for the condensate wave function. The domain size is correctly described by a mean-field Kibble-Zurek scaling exponent. The domain walls, appearing as stable topological defects, separate areas where the condensate wave function shows opposite chirality, characterized by opposite valley Chern numbers. Contrary to the quantum valley Hall interfaces, these domain walls sustain a single unidirectional mode, because the condensate selects a single valley on each side of the wall. The non-linear wave function of the domain wall and the corresponding topological one-way currents are analytically described by a solution similar to a  Jackiw-Rebbi \cite{Jackiw1976} soliton. At longer times, the system evolves towards the formation of a single valley-polarized domain, with a valley polarization chosen randomly in each experiment.

We consider a patterned microcavity under non-resonant pumping. The pattern forms a staggered honeycomb lattice with different localization energies on the $A$ and $B$ sites and a trivial gap at the Dirac point. The patterning of honeycomb and other lattices is now well-established \cite{Jacqmin2014,klembt2017polariton,Whittaker2018,Real2020}. The Dirac cones, both "straight" \cite{Jacqmin2014} and "tilted" \cite{Milicevic2019}, have been observed in polariton graphene, and the topological edge states based on the quantum anomalous Hall effect were evidenced experimentally \cite{klembt2018exciton}. Condensation has already been observed in polariton graphene at various points of the dispersion \cite{Jacqmin2014,klembt2018exciton,Suchomel2018}, depending on the experimental conditions which control the polariton relaxation and lifetime, ultimately determining the state for the condensation. In particular, condensation at the Dirac point, with a gap opened by an applied magnetic field, was observed in the work focused on the polariton topological insulator \cite{klembt2018exciton}. This is also supported by recent theoretical studies \cite{Lledo2021}.

\begin{figure}[tbp]
    \centering
    \includegraphics[width=0.99\linewidth]{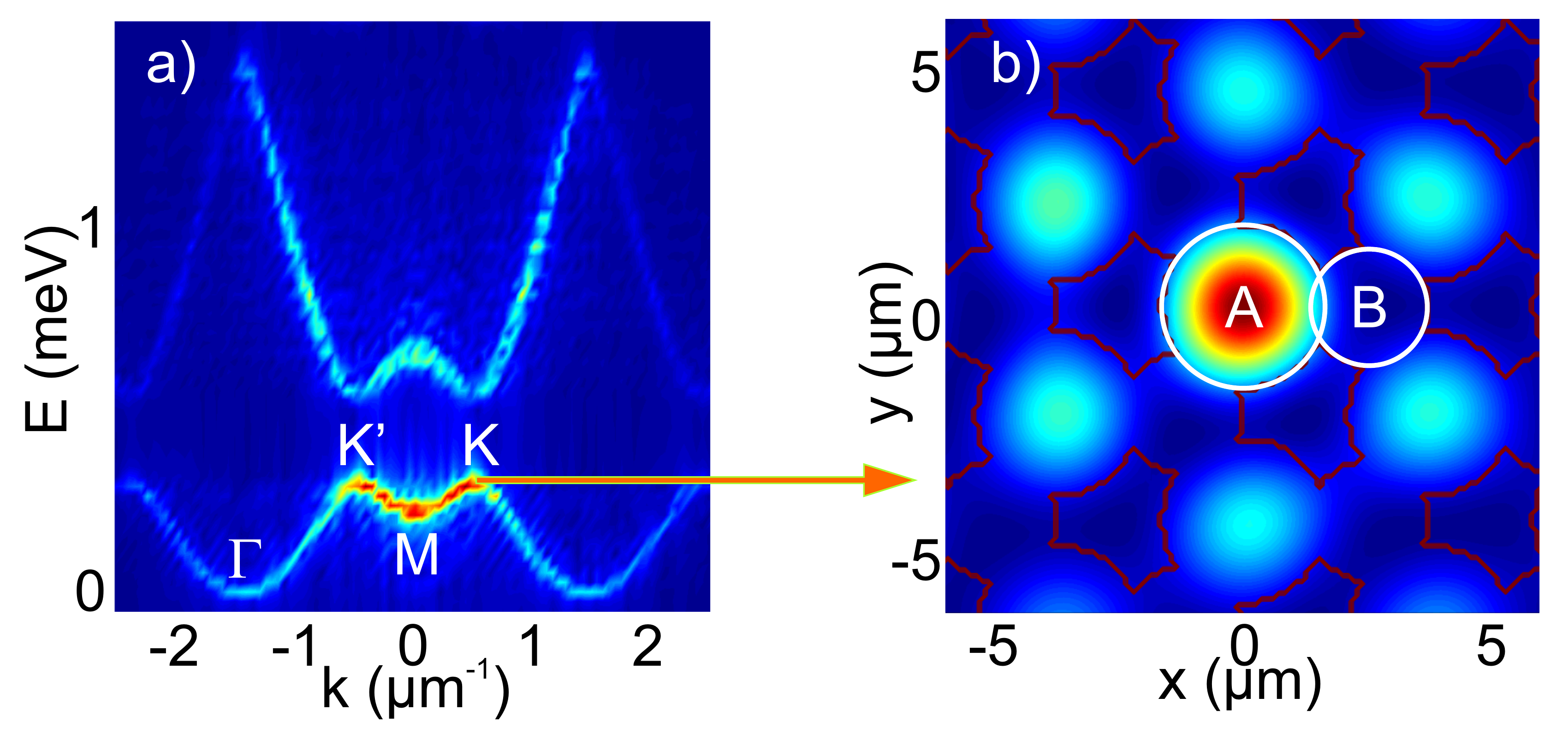}
    \caption{(a) Dispersion of the polariton graphene showing a gap at the $K$ point. (b) Spatial distribution of intensity corresponding to a single valley $K$. }
    \label{fig1}
\end{figure}

We simulate the polariton relaxation and condensation under non-resonant pumping using the Gross-Pitaevskii equation with lifetime, energy relaxation, and saturated gain
\begin{eqnarray}
    i\hbar\frac{\partial\psi}{\partial t}&=&-\left(1-i\Lambda\right)\frac{\hbar^2}{2m}\Delta\psi
    +g\left|\psi\right|^2\psi \nonumber\\
    &+&\left(U_0+U_R+i\gamma(n_{tot})-i\Gamma\right)\psi+\chi
    \label{gpe}
\end{eqnarray}
Here, $m$ is the polariton mass, $g$ is the polariton-polariton interaction constant, $U_0$ is the potential forming the staggered honeycomb lattice of polariton graphene with different site radii, $U_R$ is the  repulsive potential of the reservoir. $\gamma(\left|\psi\right|^2)=\gamma_0(n_R)\exp(-n_{tot}/n_s)$ 
is the saturated gain term with $n_{tot}=\int|\psi|^2\,dxdy$ the total particle density, $n_s$ the saturation density, and $n_R$ the exciton reservoir density. $\Gamma$ is the polariton decay time, $\chi$ is the noise describing the spontaneous scattering from the excitonic reservoir, and $\Lambda$ characterizes the efficiency of the energy relaxation \cite{Pitaevskii58}. We solve Eq.~\ref{gpe} numerically using the 3rd-order Adams method for the time derivative and a GPU-accelerated FFT for the Laplacian. We have chosen the parameters of a typical polariton graphene lattice \cite{Jacqmin2014}. This equation was already successfully used to describe polariton condensation at the $\Gamma$ point at the top of the s-band \cite{Jacqmin2014} (negative mass states) and to study theoretically the KZM at the bottom of the band \cite{GKZM}. The state where the condensate forms depends on the condensation parameters such as the lifetime of the states and on the energy relaxation efficiency \cite{Kasprzak2008,Levrat2010,Feng2013}. The latter can be controlled via the detuning (determining the excitonic fraction) and via the spot size \cite{Wertz2009,Jamadi2019}.

We begin by examining the properties of the linear states of the lattice. We take $\Lambda=g=\gamma=0$ in \eqref{gpe}. We consider a narrow (both in space and time) Gaussian wavepacket as an initial condition and apply Fourier transforms to the solution $\psi(\bm{r},t)$ in order to obtain the dispersion $|\psi(\bm{k},E)|^2$. Fig.~\ref{fig1}(a) shows a cut of the dispersion centered at the $M$ point, with a gap visible at the $K$ points. The gap size is controlled by the difference of the radius of the sites $A$ and $B$. Fig.~\ref{fig1}(b) shows the real space image of the confinement potential contour (black lines) together with the particle density $|\psi_K(x,y)|^2$ of the eigenstate at the $K$ point at the top of the lowest energy band. The two sites $A$ and $B$ are marked with white lines. The central site has a higher intensity due to the way the Eq.~\eqref{gpe} is solved numerically. For the valley extrema at the top of the valence band, the particle density on the $B$ sites is much smaller than on the $A$ sites because of the staggering.

\begin{figure}[tbp]
    \centering
    \includegraphics[width=0.99\linewidth]{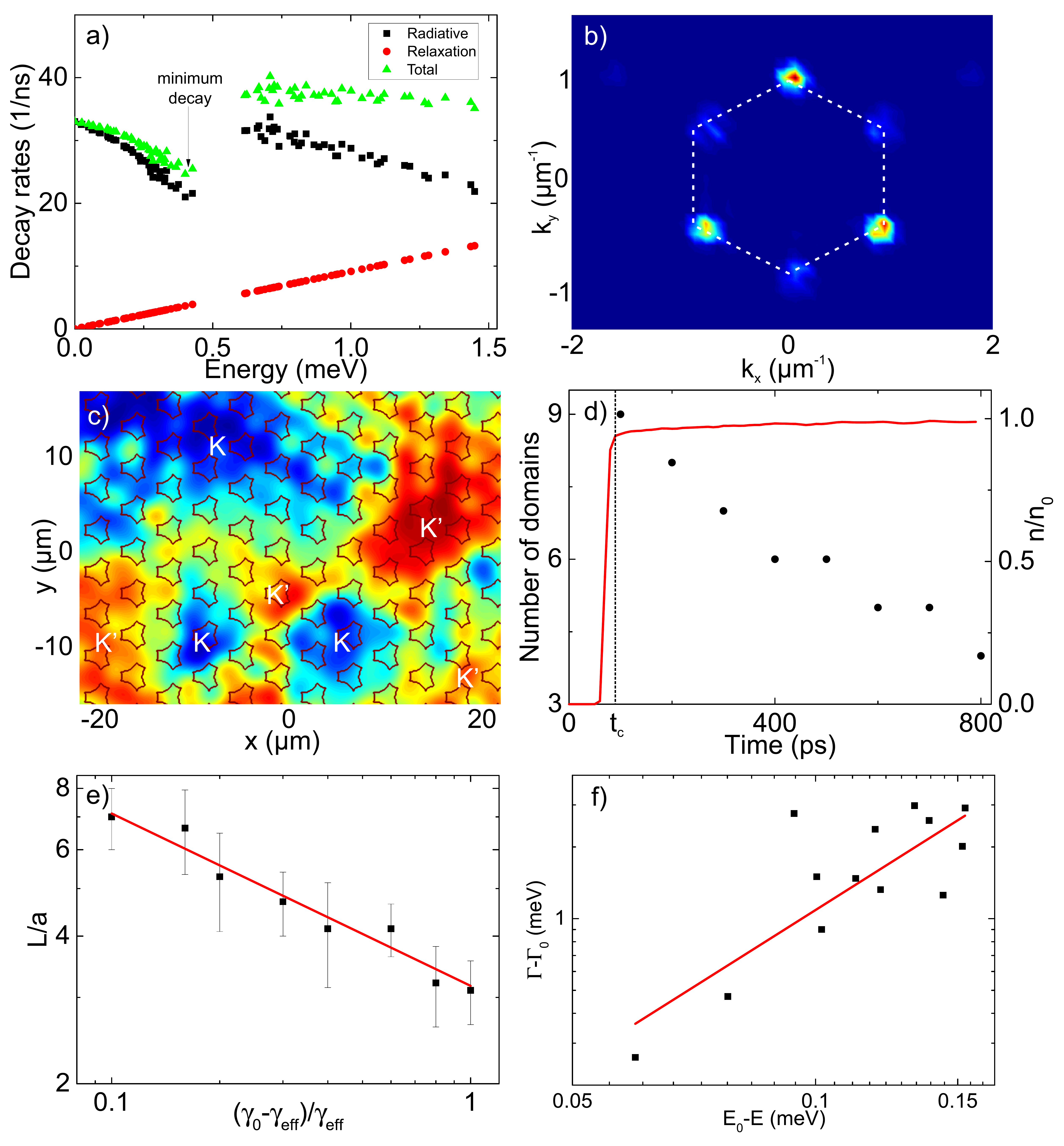}
    \caption{a) Decay rates in polariton graphene (black - radiative, red - relaxation, green - total); b) Condensed state in the reciprocal space exhibiting high intensity at the $K$ points of the Brillouin zone; c) Valley-polarized domains in real space; d) Number of domains and total particle density as functions of time; e) The domain size scaling with quench parameter (black dots with error bars) with a power law fit (red line); f) Scaling of the decay rate at the band edge (black dots) with a power law fit (red line).}
    \label{fig2}
\end{figure}

We now study the polariton relaxation and condensation in this lattice. Figure~\ref{fig2}(a) shows the decay rates calculated for the linear eigenstates of different energies (averaging over the iso-energy line in $k$-space). The radiative decay entering into Eq.~\eqref{gpe} reads $\Gamma=\Gamma_0+\Gamma_E$ where $\Gamma_0$ is approximately energy-independent and related to the losses through the cavity mirrors. $\Gamma_E$ (black points) is energy dependent  and is proportional to the intensity of the field at the surfaces of the pillars:
\begin{equation}
    \Gamma_E=\alpha\oint\limits_{surf} \left|\psi_i\left(x,y\right)\right|^2 d\ell
\end{equation}
where $d\ell$ is a line element along the etched boundary representation in the 2D model. 
The related losses are due to the presence of disorder  \cite{Jacqmin2014,Milic2018} and the suppression of radiative emission by destructive interference \cite{Aleiner2012}. 
The anti-symmetric states located at the top of bands have a smaller decay via these surface losses. The other sources of decay are scattering thermalization processes, taken into account by the $\Lambda$ coefficient in Eq.~\eqref{gpe} and which scale linearly with the energy of the states \cite{Pitaevskii58,GKZM} (red points). The sum of both rates is plotted in green: it exhibits a minimum at the top of the first band  favoring condensation at the degenerate $K$ and $K'$ points.

To simulate condensation, we consider a stationary spatial exciton distribution $n_R(x,y)=\mathrm{const}$, creating both a gain $\gamma_r$ and an interaction profile $U_R$. We then solve Eq.~\eqref{gpe} versus time. We consider a spatially homogeneous pumping and attractive polariton-polariton interactions ($g<0$) in order to create a spatially homogeneous condensate density. Because of the negative mass $m_{eff}<0$, these attractive interactions become effectively repulsive. An attractive polariton-polariton interaction can show up as a result of the interplay between the reservoir and condensate dynamics \cite{Baboux2018}. Another possibility is the polariton analogue of the Feshbach resonance \cite{Vladimirova2010,takemura2014polaritonic} related to the presence of a bi-exciton resonance, which can be realized by changing the exciton-photon detuning. (See \cite{suppl} for the usual case of repulsive interactions.) 

Condensation occurs in the $K$ and $K'$ states at the top of the valence band, because these states show the longest lifetime (see Fig.~\ref{fig2}(a)). The momentum space distribution of the condensate immediately after its formation ($t\approx t_c$) is shown in Fig.~\ref{fig2}(b), confirming the condensation at the Dirac points. Phase fluctuations present during the dynamical condensation process provoke local symmetry breaking. This leads at short times $t=t_c$ to the formation of valley-polarized domains separated by domain walls (Fig.~\ref{fig2}(c)). At longer times $t>t_c$, the valley-polarized domains change size (smaller domains shrink), to finally form a single valley-polarized domain. This is shown in Fig.~\ref{fig2}(d), showing the time evolution of the mean condensate density and of the number of domains. The final valley polarization achieved ($K$ or $K'$) is randomly chosen for each experiment. This long-time result is similar to the one recently found in \cite{Lledo2021}.  

The formation of domains in second-order phase transitions is described by the Kibble-Zurek mechanism, where the quench time is controlled by the normalized pumping density $(\gamma_0-\gamma_{eff})/\gamma_{eff}$, where $\gamma_0$ is the reservoir gain controlled by the pumping and $\gamma_{eff}$ is its critical value, below which the condensation does not occur  \cite{Solnyshkov2016,GKZM}. The dependence of the size of KZM valley-polarized domains versus the quenching parameter \cite{Zurek1996} is  shown in Fig.~\ref{fig2}(e). It follows a power law decay with a scaling exponent $\eta=-0.34\pm 0.03$. In the mean-field approximation the KZM scaling exponent for the domain size reads: 
\begin{equation}
\eta=-(D-d)\frac{\nu}{1+z\nu}
\end{equation}
where $D=2$ and $d=1$ are the space and domain wall dimensionalities. $z\nu$ is the dynamical scaling exponent, given by the energy dependence of the total decay rate at the band edge (Fig.~\ref{fig2}(a)). Figure~\ref{fig2}(f) shows the decay rate as a function of energy, with the band edge taken as a zero reference. Indeed, in a stationary configuration this decay is exactly compensated by the gain, giving a zero net decay for the condensate. All other states exhibit a stronger decay rate. The energy is measured from the band edge towards the bottom (negative mass states). The scaling of the decay rate appears to be $2.1\pm 0.3$, consistent with a dynamical scaling exponent $z\nu =2$. Together with the critical exponent $\nu=1$ found previously for a honeycomb lattice \cite{GKZM} and appearing due to the linear dispersion of the Dirac cone, this gives a mean-field KZM scaling exponent $\eta=-1/3$, in excellent agreement with numerical simulations (Fig.~\ref{fig2}(e)).

\begin{figure}[tbp]
    \centering
    \includegraphics[width=0.99\linewidth]{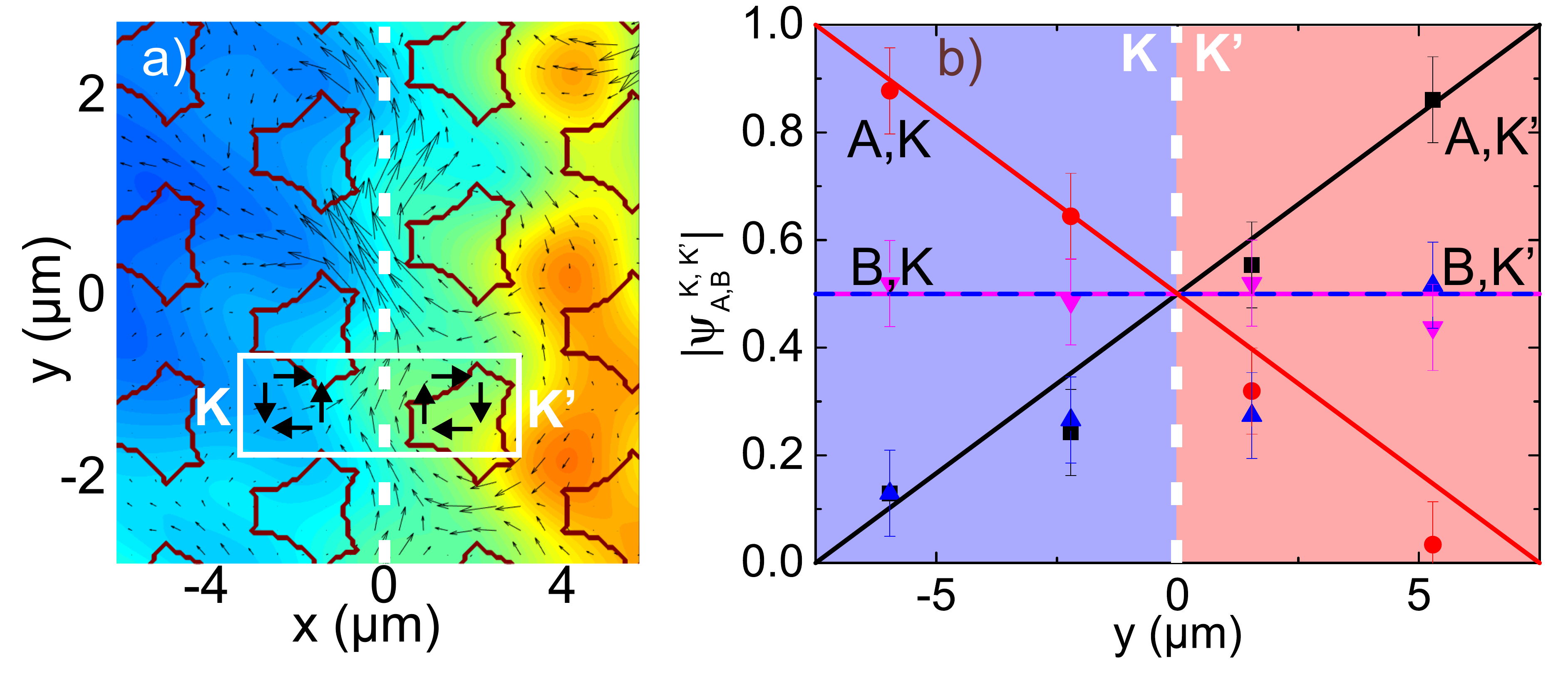}
    \caption{(a) Two valley-polarized domains (false color shows the valley polarization) with a boundary (white dashed line) exhibiting a localized one-way current (black arrows). The effective fields of the two occupied valleys with opposite windings are shown with black arrows in the white rectangle. (b) The wave function components across the interface: numerics (dots) and analytical ansatz~\eqref{solution} (lines).  }
    \label{fig3}
\end{figure}

We now study the domain walls, where the continuity of the condensate wave function needs to be ensured. Figure~\ref{fig3}(a) shows the quantum-mechanical current as a function of coordinate (arrows) together with valley polarization (false color). The domain wall is marked with a white dashed line. No net current is flowing through the sites within the valleys. On the contrary, the domain wall clearly carries a net current oriented upwards.
 
This one-way interface current has deep topological roots, being qualitatively similar to the well-known chiral Jackiw-Rebbi \cite{Jackiw1976} interface state between two regions described by Dirac equations with opposite masses \cite{Hasan2010}. However, there are important differences between the ideal Jackiw-Rebbi case and the present configuration. In our case the Hamiltonian on both sides of the domain wall is the same (and describes both valleys), it is the wave function, which is the solution of the non-linear Dirac equation, which changes its valley polarization across the domain wall. The domains and the domain walls are therefore dynamical objects capable of evolution. However, the origin of chirality is the same in both cases: the opposite winding (Berry curvature) of the two valleys imposes the pseudospin texture and the group velocity near the interface. The spatial image in Fig.~\ref{fig3}(a) shows the interface and the distribution of the effective field in the two valleys, exhibiting opposite windings (the arrows inside the white rectangle).

A system with two valleys can be described with a $4\times 4$ block-diagonal Hamiltonian composed of two 2D Dirac Hamiltonians with opposite winding. Including the on-site interactions (valley-isotropic, but site-dependent), the explicit stationary non-linear Dirac equation reads:
\begin{widetext}
\small
\begin{equation}
\left( {\begin{array}{*{20}{c}}
{\Delta  + g{{\left| {\psi _A^K} \right|}^2} + g{{\left| {\psi _A^{K'}} \right|}^2}}&{\hbar c\left( {- \frac{\partial }{{\partial x}} -i \frac{\partial }{{\partial y}}} \right)}&0&0\\
{\hbar c\left( { +\frac{\partial }{{\partial x}} -i \frac{\partial }{{\partial y}}} \right)}&{ - \Delta  + g{{\left| {\psi _B^K} \right|}^2} + g{{\left| {\psi _B^{K'}} \right|}^2}}&0&0\\
0&0&{\Delta  + g{{\left| {\psi _A^K} \right|}^2} + g{{\left| {\psi _A^{K'}} \right|}^2}}&{\hbar c\left( { + \frac{\partial }{{\partial x}} -i \frac{\partial }{{\partial y}}} \right)}\\
0&0&{\hbar c\left( { -\frac{\partial }{{\partial x}} -i\frac{\partial }{{\partial y}}} \right)}&{ - \Delta  + g{{\left| {\psi _B^K} \right|}^2} + g{{\left| {\psi _B^{K'}} \right|}^2}}
\end{array}} \right)\left( {\begin{array}{*{20}{c}}
{\psi _A^K}\\
{\psi _B^K}\\
{\psi _A^{K'}}\\
{\psi _B^{K'}}
\end{array}} \right) = E\left( {\begin{array}{*{20}{c}}
{\psi _A^K}\\
{\psi _B^K}\\
{\psi _A^{K'}}\\
{\psi _B^{K'}}
\end{array}} \right)
\label{bigDirac}
\end{equation}
\end{widetext}
where $\Delta$ is the mass term due to the graphene staggering, identical for both valleys, and $g$ is the on-site interaction constant. Far from the interface, the valley-polarized condensate wave functions are given by $\psi_{+\infty}=(1,0,0,0)^T$ and $\psi_{-\infty}=(0,0,1,0)^T$. These boundary conditions are induced by the spontaneous  symmetry breaking during the condensation. Exactly at the interface, all 4 components have to be non-zero in order to satisfy Eq.~\eqref{bigDirac}. 

We find that the solution of the non-linear Dirac equation with two valley-polarized regions and a domain wall between them can be written as
\begin{equation}
    \left| \psi  \right\rangle \approx  \left( {\begin{array}{*{20}{c}}
{\frac{1}{2} + ax}\\
{\frac{1}{2} }\\
{\frac{1}{2} - ax}\\
{\frac{1}{2} }
\end{array}} \right)e^{ik_y y}
\label{solution}
\end{equation}
This expression is indeed a solution (see \cite{suppl} for details) of the Dirac equation~\eqref{bigDirac}, valid to the first order in $x$ and $k_y$ near $x=0$, with $a=\Delta/\hbar c$ (the inverse Compton wavelength, recently associated with the quantum metric~\cite{Leblanc2021}) and $E=\Delta+g/2+\hbar c k_y$ (with $\Delta<0$ and $g<0$). 
This solution is shown in Fig.~\ref{fig3}(b) with lines (whose color corresponds to the wave function components), together with the points, extracted from the wave function of the numerical experiment shown in Fig.~\ref{fig3}(a). The good agreement of the analytical ansatz with the numerical solution confirms the validity of the former.

From $E(k_y)$, we obtain the group velocity $+c$ along the $y$ axis. Its well-defined sign indicates a one-way state. The opposite direction is forbidden because of the pseudospin orientation at the interface. Any other pseudospin orientation is not a solution of \eqref{bigDirac} for this interface configuration, just as in the Jackiw-Rebbi model. The opposite behavior of the 1st and 3rd components across the domain wall is possible thanks to the opposite valleys winding (opposite sign of $\partial/\partial x$ in \eqref{bigDirac}). At the same time, the constant sign of $\partial/\partial y$ provides a single direction for the current along the interface. We thus conclude that the chiral localized current along the domain wall solution of the non-linear Dirac equation has the same origin as the chiral interface state in linear Dirac equation with inverted mass: the opposite topology on both sides of the interface. Its localization length is determined by the gap size $l=1/a=\hbar c/\Delta$.

To conclude, we have shown that condensation can occur at the Dirac points in staggered polariton graphene. Under homogeneous pumping with repulsive effective interactions, valley-polarized domains form via the Kibble-Zurek mechanism. Stable domain walls between such domains are carrying topological one-way currents.

\begin{acknowledgments}
We acknowledge the support of the European Union's Horizon 2020 program, through a FET Open research and innovation action under the grant agreement No. 964770 (TopoLight), project ANR Labex GaNEXT (ANR-11-LABX-0014), and of the ANR program "Investissements d'Avenir" through the IDEX-ISITE initiative 16-IDEX-0001 (CAP 20-25). 
\end{acknowledgments}

\bibliography{biblio}

\begin{thebibliography}{62}%
\makeatletter
\providecommand \@ifxundefined [1]{%
 \@ifx{#1\undefined}
}%
\providecommand \@ifnum [1]{%
 \ifnum #1\expandafter \@firstoftwo
 \else \expandafter \@secondoftwo
 \fi
}%
\providecommand \@ifx [1]{%
 \ifx #1\expandafter \@firstoftwo
 \else \expandafter \@secondoftwo
 \fi
}%
\providecommand \natexlab [1]{#1}%
\providecommand \enquote  [1]{``#1''}%
\providecommand \bibnamefont  [1]{#1}%
\providecommand \bibfnamefont [1]{#1}%
\providecommand \citenamefont [1]{#1}%
\providecommand \href@noop [0]{\@secondoftwo}%
\providecommand \href [0]{\begingroup \@sanitize@url \@href}%
\providecommand \@href[1]{\@@startlink{#1}\@@href}%
\providecommand \@@href[1]{\endgroup#1\@@endlink}%
\providecommand \@sanitize@url [0]{\catcode `\\12\catcode `\$12\catcode
  `\&12\catcode `\#12\catcode `\^12\catcode `\_12\catcode `\%12\relax}%
\providecommand \@@startlink[1]{}%
\providecommand \@@endlink[0]{}%
\providecommand \url  [0]{\begingroup\@sanitize@url \@url }%
\providecommand \@url [1]{\endgroup\@href {#1}{\urlprefix }}%
\providecommand \urlprefix  [0]{URL }%
\providecommand \Eprint [0]{\href }%
\providecommand \doibase [0]{https://doi.org/}%
\providecommand \selectlanguage [0]{\@gobble}%
\providecommand \bibinfo  [0]{\@secondoftwo}%
\providecommand \bibfield  [0]{\@secondoftwo}%
\providecommand \translation [1]{[#1]}%
\providecommand \BibitemOpen [0]{}%
\providecommand \bibitemStop [0]{}%
\providecommand \bibitemNoStop [0]{.\EOS\space}%
\providecommand \EOS [0]{\spacefactor3000\relax}%
\providecommand \BibitemShut  [1]{\csname bibitem#1\endcsname}%
\let\auto@bib@innerbib\@empty
\bibitem [{\citenamefont {Vergniory}\ \emph {et~al.}(2019)\citenamefont
  {Vergniory}, \citenamefont {Elcoro}, \citenamefont {Felser}, \citenamefont
  {Regnault}, \citenamefont {Bernevig},\ and\ \citenamefont
  {Wang}}]{Vergniory2019}%
  \BibitemOpen
  \bibfield  {author} {\bibinfo {author} {\bibfnamefont {M.~G.}\ \bibnamefont
  {Vergniory}}, \bibinfo {author} {\bibfnamefont {L.}~\bibnamefont {Elcoro}},
  \bibinfo {author} {\bibfnamefont {C.}~\bibnamefont {Felser}}, \bibinfo
  {author} {\bibfnamefont {N.}~\bibnamefont {Regnault}}, \bibinfo {author}
  {\bibfnamefont {B.~A.}\ \bibnamefont {Bernevig}},\ and\ \bibinfo {author}
  {\bibfnamefont {Z.}~\bibnamefont {Wang}},\ }\bibfield  {title} {\bibinfo
  {title} {A complete catalogue of high-quality topological materials},\
  }\href@noop {} {\bibfield  {journal} {\bibinfo  {journal} {Nature}\ }\textbf
  {\bibinfo {volume} {566}},\ \bibinfo {pages} {480} (\bibinfo {year}
  {2019})}\BibitemShut {NoStop}%
\bibitem [{\citenamefont {Das~Sarma}\ \emph {et~al.}(2005)\citenamefont
  {Das~Sarma}, \citenamefont {Freedman},\ and\ \citenamefont
  {Nayak}}]{DasSarma2005}%
  \BibitemOpen
  \bibfield  {author} {\bibinfo {author} {\bibfnamefont {S.}~\bibnamefont
  {Das~Sarma}}, \bibinfo {author} {\bibfnamefont {M.}~\bibnamefont
  {Freedman}},\ and\ \bibinfo {author} {\bibfnamefont {C.}~\bibnamefont
  {Nayak}},\ }\bibfield  {title} {\bibinfo {title} {Topologically protected
  qubits from a possible non-abelian fractional quantum hall state},\ }\href
  {https://doi.org/10.1103/PhysRevLett.94.166802} {\bibfield  {journal}
  {\bibinfo  {journal} {Phys. Rev. Lett.}\ }\textbf {\bibinfo {volume} {94}},\
  \bibinfo {pages} {166802} (\bibinfo {year} {2005})}\BibitemShut {NoStop}%
\bibitem [{\citenamefont {Gladchenko}\ \emph {et~al.}(2009)\citenamefont
  {Gladchenko}, \citenamefont {Olaya}, \citenamefont {Dupont-Ferrier},
  \citenamefont {Doucot}, \citenamefont {Ioffe},\ and\ \citenamefont
  {Gershenson}}]{Gladchenko2009}%
  \BibitemOpen
  \bibfield  {author} {\bibinfo {author} {\bibfnamefont {S.}~\bibnamefont
  {Gladchenko}}, \bibinfo {author} {\bibfnamefont {D.}~\bibnamefont {Olaya}},
  \bibinfo {author} {\bibfnamefont {E.}~\bibnamefont {Dupont-Ferrier}},
  \bibinfo {author} {\bibfnamefont {B.}~\bibnamefont {Doucot}}, \bibinfo
  {author} {\bibfnamefont {L.~B.}\ \bibnamefont {Ioffe}},\ and\ \bibinfo
  {author} {\bibfnamefont {M.~E.}\ \bibnamefont {Gershenson}},\ }\bibfield
  {title} {\bibinfo {title} {Superconducting nanocircuits for topologically
  protected qubits},\ }\href@noop {} {\bibfield  {journal} {\bibinfo  {journal}
  {Nat. Phys.}\ }\textbf {\bibinfo {volume} {5}},\ \bibinfo {pages} {48}
  (\bibinfo {year} {2009})}\BibitemShut {NoStop}%
\bibitem [{\citenamefont {Bahari}\ \emph {et~al.}(2017)\citenamefont {Bahari},
  \citenamefont {Ndao}, \citenamefont {Vallini}, \citenamefont {{El Amili}},
  \citenamefont {Fainman},\ and\ \citenamefont {Kant{\'{e}}}}]{Bahari2017}%
  \BibitemOpen
  \bibfield  {author} {\bibinfo {author} {\bibfnamefont {B.}~\bibnamefont
  {Bahari}}, \bibinfo {author} {\bibfnamefont {A.}~\bibnamefont {Ndao}},
  \bibinfo {author} {\bibfnamefont {F.}~\bibnamefont {Vallini}}, \bibinfo
  {author} {\bibfnamefont {A.}~\bibnamefont {{El Amili}}}, \bibinfo {author}
  {\bibfnamefont {Y.}~\bibnamefont {Fainman}},\ and\ \bibinfo {author}
  {\bibfnamefont {B.}~\bibnamefont {Kant{\'{e}}}},\ }\bibfield  {title}
  {\bibinfo {title} {{Nonreciprocal lasing in topological cavities of arbitrary
  geometries}},\ }\href {https://doi.org/10.1126/science.aao4551} {\bibfield
  {journal} {\bibinfo  {journal} {Science (New York, N.Y.)}\ }\textbf {\bibinfo
  {volume} {358}},\ \bibinfo {pages} {636} (\bibinfo {year}
  {2017})}\BibitemShut {NoStop}%
\bibitem [{\citenamefont {Bandres}\ \emph {et~al.}(2018)\citenamefont
  {Bandres}, \citenamefont {Wittek}, \citenamefont {Harari}, \citenamefont
  {Parto}, \citenamefont {Ren}, \citenamefont {Segev}, \citenamefont
  {Christodoulides},\ and\ \citenamefont {Khajavikhan}}]{Bandres2018}%
  \BibitemOpen
  \bibfield  {author} {\bibinfo {author} {\bibfnamefont {M.~A.}\ \bibnamefont
  {Bandres}}, \bibinfo {author} {\bibfnamefont {S.}~\bibnamefont {Wittek}},
  \bibinfo {author} {\bibfnamefont {G.}~\bibnamefont {Harari}}, \bibinfo
  {author} {\bibfnamefont {M.}~\bibnamefont {Parto}}, \bibinfo {author}
  {\bibfnamefont {J.}~\bibnamefont {Ren}}, \bibinfo {author} {\bibfnamefont
  {M.}~\bibnamefont {Segev}}, \bibinfo {author} {\bibfnamefont {D.~N.}\
  \bibnamefont {Christodoulides}},\ and\ \bibinfo {author} {\bibfnamefont
  {M.}~\bibnamefont {Khajavikhan}},\ }\bibfield  {title} {\bibinfo {title}
  {{Topological insulator laser: Experiments}},\ }\href
  {http://science.sciencemag.org/content/359/6381/eaar4005.abstract} {\bibfield
   {journal} {\bibinfo  {journal} {Science (New York, N.Y.)}\ }\textbf
  {\bibinfo {volume} {359}},\ \bibinfo {pages} {aar4005} (\bibinfo {year}
  {2018})}\BibitemShut {NoStop}%
\bibitem [{\citenamefont {Solnyshkov}\ \emph {et~al.}(2018)\citenamefont
  {Solnyshkov}, \citenamefont {Bleu},\ and\ \citenamefont
  {Malpuech}}]{Solnyshkov2018}%
  \BibitemOpen
  \bibfield  {author} {\bibinfo {author} {\bibfnamefont {D.~D.}\ \bibnamefont
  {Solnyshkov}}, \bibinfo {author} {\bibfnamefont {O.}~\bibnamefont {Bleu}},\
  and\ \bibinfo {author} {\bibfnamefont {G.}~\bibnamefont {Malpuech}},\
  }\bibfield  {title} {\bibinfo {title} {{Topological optical isolator based on
  polariton graphene}},\ }\href {https://doi.org/10.1063/1.5018902} {\bibfield
  {journal} {\bibinfo  {journal} {Applied Physics Letters}\ }\textbf {\bibinfo
  {volume} {112}},\ \bibinfo {pages} {31106} (\bibinfo {year}
  {2018})}\BibitemShut {NoStop}%
\bibitem [{\citenamefont {Karki}\ \emph {et~al.}(2019)\citenamefont {Karki},
  \citenamefont {El-Ganainy},\ and\ \citenamefont {Levy}}]{Karki2019}%
  \BibitemOpen
  \bibfield  {author} {\bibinfo {author} {\bibfnamefont {D.}~\bibnamefont
  {Karki}}, \bibinfo {author} {\bibfnamefont {R.}~\bibnamefont {El-Ganainy}},\
  and\ \bibinfo {author} {\bibfnamefont {M.}~\bibnamefont {Levy}},\ }\bibfield
  {title} {\bibinfo {title} {Toward high-performing topological edge-state
  optical isolators},\ }\href
  {https://doi.org/10.1103/PhysRevApplied.11.034045} {\bibfield  {journal}
  {\bibinfo  {journal} {Phys. Rev. Applied}\ }\textbf {\bibinfo {volume}
  {11}},\ \bibinfo {pages} {034045} (\bibinfo {year} {2019})}\BibitemShut
  {NoStop}%
\bibitem [{\citenamefont {Novoselov}\ \emph {et~al.}(2005)\citenamefont
  {Novoselov}, \citenamefont {Jiang}, \citenamefont {Schedin}, \citenamefont
  {Booth}, \citenamefont {Khotkevich}, \citenamefont {Morozov},\ and\
  \citenamefont {Geim}}]{novoselov2005two}%
  \BibitemOpen
  \bibfield  {author} {\bibinfo {author} {\bibfnamefont {K.~S.}\ \bibnamefont
  {Novoselov}}, \bibinfo {author} {\bibfnamefont {D.}~\bibnamefont {Jiang}},
  \bibinfo {author} {\bibfnamefont {F.}~\bibnamefont {Schedin}}, \bibinfo
  {author} {\bibfnamefont {T.}~\bibnamefont {Booth}}, \bibinfo {author}
  {\bibfnamefont {V.}~\bibnamefont {Khotkevich}}, \bibinfo {author}
  {\bibfnamefont {S.}~\bibnamefont {Morozov}},\ and\ \bibinfo {author}
  {\bibfnamefont {A.~K.}\ \bibnamefont {Geim}},\ }\bibfield  {title} {\bibinfo
  {title} {Two-dimensional atomic crystals},\ }\href@noop {} {\bibfield
  {journal} {\bibinfo  {journal} {Proceedings of the National Academy of
  Sciences}\ }\textbf {\bibinfo {volume} {102}},\ \bibinfo {pages} {10451}
  (\bibinfo {year} {2005})}\BibitemShut {NoStop}%
\bibitem [{\citenamefont {Novoselov}\ \emph {et~al.}(2016)\citenamefont
  {Novoselov}, \citenamefont {Mishchenko}, \citenamefont {Carvalho},\ and\
  \citenamefont {Neto}}]{novoselov20162d}%
  \BibitemOpen
  \bibfield  {author} {\bibinfo {author} {\bibfnamefont {K.}~\bibnamefont
  {Novoselov}}, \bibinfo {author} {\bibfnamefont {o.~A.}\ \bibnamefont
  {Mishchenko}}, \bibinfo {author} {\bibfnamefont {o.~A.}\ \bibnamefont
  {Carvalho}},\ and\ \bibinfo {author} {\bibfnamefont {A.~C.}\ \bibnamefont
  {Neto}},\ }\bibfield  {title} {\bibinfo {title} {2d materials and van der
  waals heterostructures},\ }\href@noop {} {\bibfield  {journal} {\bibinfo
  {journal} {Science}\ }\textbf {\bibinfo {volume} {353}} (\bibinfo {year}
  {2016})}\BibitemShut {NoStop}%
\bibitem [{\citenamefont {Noh}\ \emph {et~al.}(2018)\citenamefont {Noh},
  \citenamefont {Huang}, \citenamefont {Chen},\ and\ \citenamefont
  {Rechtsman}}]{Noh2018}%
  \BibitemOpen
  \bibfield  {author} {\bibinfo {author} {\bibfnamefont {J.}~\bibnamefont
  {Noh}}, \bibinfo {author} {\bibfnamefont {S.}~\bibnamefont {Huang}}, \bibinfo
  {author} {\bibfnamefont {K.~P.}\ \bibnamefont {Chen}},\ and\ \bibinfo
  {author} {\bibfnamefont {M.~C.}\ \bibnamefont {Rechtsman}},\ }\bibfield
  {title} {\bibinfo {title} {Observation of photonic topological valley hall
  edge states},\ }\href {https://doi.org/10.1103/PhysRevLett.120.063902}
  {\bibfield  {journal} {\bibinfo  {journal} {Phys. Rev. Lett.}\ }\textbf
  {\bibinfo {volume} {120}},\ \bibinfo {pages} {063902} (\bibinfo {year}
  {2018})}\BibitemShut {NoStop}%
\bibitem [{\citenamefont {Xiao}\ \emph {et~al.}(2007)\citenamefont {Xiao},
  \citenamefont {Yao},\ and\ \citenamefont {Niu}}]{Niu2007}%
  \BibitemOpen
  \bibfield  {author} {\bibinfo {author} {\bibfnamefont {D.}~\bibnamefont
  {Xiao}}, \bibinfo {author} {\bibfnamefont {W.}~\bibnamefont {Yao}},\ and\
  \bibinfo {author} {\bibfnamefont {Q.}~\bibnamefont {Niu}},\ }\bibfield
  {title} {\bibinfo {title} {Valley-contrasting physics in graphene: Magnetic
  moment and topological transport},\ }\href
  {https://doi.org/10.1103/PhysRevLett.99.236809} {\bibfield  {journal}
  {\bibinfo  {journal} {Phys. Rev. Lett.}\ }\textbf {\bibinfo {volume} {99}},\
  \bibinfo {pages} {236809} (\bibinfo {year} {2007})}\BibitemShut {NoStop}%
\bibitem [{\citenamefont {Yao}\ \emph {et~al.}(2009)\citenamefont {Yao},
  \citenamefont {Yang},\ and\ \citenamefont {Niu}}]{Yao2009}%
  \BibitemOpen
  \bibfield  {author} {\bibinfo {author} {\bibfnamefont {W.}~\bibnamefont
  {Yao}}, \bibinfo {author} {\bibfnamefont {S.~A.}\ \bibnamefont {Yang}},\ and\
  \bibinfo {author} {\bibfnamefont {Q.}~\bibnamefont {Niu}},\ }\bibfield
  {title} {\bibinfo {title} {Edge states in graphene: From gapped flat-band to
  gapless chiral modes},\ }\href
  {https://doi.org/10.1103/PhysRevLett.102.096801} {\bibfield  {journal}
  {\bibinfo  {journal} {Phys. Rev. Lett.}\ }\textbf {\bibinfo {volume} {102}},\
  \bibinfo {pages} {096801} (\bibinfo {year} {2009})}\BibitemShut {NoStop}%
\bibitem [{\citenamefont {Ma}\ \emph {et~al.}(2015)\citenamefont {Ma},
  \citenamefont {Khanikaev}, \citenamefont {Mousavi},\ and\ \citenamefont
  {Shvets}}]{Ma2015}%
  \BibitemOpen
  \bibfield  {author} {\bibinfo {author} {\bibfnamefont {T.}~\bibnamefont
  {Ma}}, \bibinfo {author} {\bibfnamefont {A.~B.}\ \bibnamefont {Khanikaev}},
  \bibinfo {author} {\bibfnamefont {S.~H.}\ \bibnamefont {Mousavi}},\ and\
  \bibinfo {author} {\bibfnamefont {G.}~\bibnamefont {Shvets}},\ }\bibfield
  {title} {\bibinfo {title} {Guiding electromagnetic waves around sharp
  corners: Topologically protected photonic transport in metawaveguides},\
  }\href {https://doi.org/10.1103/PhysRevLett.114.127401} {\bibfield  {journal}
  {\bibinfo  {journal} {Phys. Rev. Lett.}\ }\textbf {\bibinfo {volume} {114}},\
  \bibinfo {pages} {127401} (\bibinfo {year} {2015})}\BibitemShut {NoStop}%
\bibitem [{\citenamefont {Ju}\ \emph {et~al.}(2015)\citenamefont {Ju},
  \citenamefont {Shi}, \citenamefont {Nair}, \citenamefont {Lv}, \citenamefont
  {Jin}, \citenamefont {Velasco}, \citenamefont {Ojeda-Aristizabal},
  \citenamefont {Bechtel}, \citenamefont {Martin}, \citenamefont {Zettl} \emph
  {et~al.}}]{ju2015topological}%
  \BibitemOpen
  \bibfield  {author} {\bibinfo {author} {\bibfnamefont {L.}~\bibnamefont
  {Ju}}, \bibinfo {author} {\bibfnamefont {Z.}~\bibnamefont {Shi}}, \bibinfo
  {author} {\bibfnamefont {N.}~\bibnamefont {Nair}}, \bibinfo {author}
  {\bibfnamefont {Y.}~\bibnamefont {Lv}}, \bibinfo {author} {\bibfnamefont
  {C.}~\bibnamefont {Jin}}, \bibinfo {author} {\bibfnamefont {J.}~\bibnamefont
  {Velasco}}, \bibinfo {author} {\bibfnamefont {C.}~\bibnamefont
  {Ojeda-Aristizabal}}, \bibinfo {author} {\bibfnamefont {H.~A.}\ \bibnamefont
  {Bechtel}}, \bibinfo {author} {\bibfnamefont {M.~C.}\ \bibnamefont {Martin}},
  \bibinfo {author} {\bibfnamefont {A.}~\bibnamefont {Zettl}}, \emph {et~al.},\
  }\bibfield  {title} {\bibinfo {title} {Topological valley transport at
  bilayer graphene domain walls},\ }\href@noop {} {\bibfield  {journal}
  {\bibinfo  {journal} {Nature}\ }\textbf {\bibinfo {volume} {520}},\ \bibinfo
  {pages} {650} (\bibinfo {year} {2015})}\BibitemShut {NoStop}%
\bibitem [{\citenamefont {Xu}\ \emph {et~al.}(2015)\citenamefont {Xu},
  \citenamefont {Belopolski}, \citenamefont {Alidoust}, \citenamefont
  {Neupane}, \citenamefont {Bian}, \citenamefont {Zhang}, \citenamefont
  {Sankar}, \citenamefont {Chang}, \citenamefont {Yuan}, \citenamefont {Lee},
  \citenamefont {Huang}, \citenamefont {Zheng}, \citenamefont {Ma},
  \citenamefont {Sanchez}, \citenamefont {Wang}, \citenamefont {Bansil},
  \citenamefont {Chou}, \citenamefont {Shibayev}, \citenamefont {Lin},
  \citenamefont {Jia},\ and\ \citenamefont {Hasan}}]{xu2015discovery}%
  \BibitemOpen
  \bibfield  {author} {\bibinfo {author} {\bibfnamefont {S.-Y.}\ \bibnamefont
  {Xu}}, \bibinfo {author} {\bibfnamefont {I.}~\bibnamefont {Belopolski}},
  \bibinfo {author} {\bibfnamefont {N.}~\bibnamefont {Alidoust}}, \bibinfo
  {author} {\bibfnamefont {M.}~\bibnamefont {Neupane}}, \bibinfo {author}
  {\bibfnamefont {G.}~\bibnamefont {Bian}}, \bibinfo {author} {\bibfnamefont
  {C.}~\bibnamefont {Zhang}}, \bibinfo {author} {\bibfnamefont
  {R.}~\bibnamefont {Sankar}}, \bibinfo {author} {\bibfnamefont
  {G.}~\bibnamefont {Chang}}, \bibinfo {author} {\bibfnamefont
  {Z.}~\bibnamefont {Yuan}}, \bibinfo {author} {\bibfnamefont {C.-C.}\
  \bibnamefont {Lee}}, \bibinfo {author} {\bibfnamefont {S.-M.}\ \bibnamefont
  {Huang}}, \bibinfo {author} {\bibfnamefont {H.}~\bibnamefont {Zheng}},
  \bibinfo {author} {\bibfnamefont {J.}~\bibnamefont {Ma}}, \bibinfo {author}
  {\bibfnamefont {D.~S.}\ \bibnamefont {Sanchez}}, \bibinfo {author}
  {\bibfnamefont {B.}~\bibnamefont {Wang}}, \bibinfo {author} {\bibfnamefont
  {A.}~\bibnamefont {Bansil}}, \bibinfo {author} {\bibfnamefont
  {F.}~\bibnamefont {Chou}}, \bibinfo {author} {\bibfnamefont {P.}~\bibnamefont
  {Shibayev}}, \bibinfo {author} {\bibfnamefont {H.}~\bibnamefont {Lin}},
  \bibinfo {author} {\bibfnamefont {S.}~\bibnamefont {Jia}},\ and\ \bibinfo
  {author} {\bibfnamefont {M.~Z.}\ \bibnamefont {Hasan}},\ }\bibfield  {title}
  {\bibinfo {title} {Discovery of a weyl fermion semimetal and topological
  fermi arcs},\ }\href@noop {} {\bibfield  {journal} {\bibinfo  {journal}
  {Science}\ }\textbf {\bibinfo {volume} {349}},\ \bibinfo {pages} {613}
  (\bibinfo {year} {2015})}\BibitemShut {NoStop}%
\bibitem [{\citenamefont {Ashcroft}\ and\ \citenamefont
  {Mermin}(1976)}]{AshcroftMermin}%
  \BibitemOpen
  \bibfield  {author} {\bibinfo {author} {\bibfnamefont {N.~W.}\ \bibnamefont
  {Ashcroft}}\ and\ \bibinfo {author} {\bibfnamefont {N.~D.}\ \bibnamefont
  {Mermin}},\ }\href@noop {} {\emph {\bibinfo {title} {Solid State Physics}}}\
  (\bibinfo  {publisher} {Harcourt, Inc. (New York, USA)},\ \bibinfo {year}
  {1976})\BibitemShut {NoStop}%
\bibitem [{\citenamefont {Wu}\ \emph {et~al.}(2016)\citenamefont {Wu},
  \citenamefont {Zhang}, \citenamefont {Sun}, \citenamefont {Xu}, \citenamefont
  {Wang}, \citenamefont {Ji}, \citenamefont {Deng}, \citenamefont {Chen},
  \citenamefont {Liu},\ and\ \citenamefont {Pan}}]{wu2016realization}%
  \BibitemOpen
  \bibfield  {author} {\bibinfo {author} {\bibfnamefont {Z.}~\bibnamefont
  {Wu}}, \bibinfo {author} {\bibfnamefont {L.}~\bibnamefont {Zhang}}, \bibinfo
  {author} {\bibfnamefont {W.}~\bibnamefont {Sun}}, \bibinfo {author}
  {\bibfnamefont {X.-T.}\ \bibnamefont {Xu}}, \bibinfo {author} {\bibfnamefont
  {B.-Z.}\ \bibnamefont {Wang}}, \bibinfo {author} {\bibfnamefont {S.-C.}\
  \bibnamefont {Ji}}, \bibinfo {author} {\bibfnamefont {Y.}~\bibnamefont
  {Deng}}, \bibinfo {author} {\bibfnamefont {S.}~\bibnamefont {Chen}}, \bibinfo
  {author} {\bibfnamefont {X.-J.}\ \bibnamefont {Liu}},\ and\ \bibinfo {author}
  {\bibfnamefont {J.-W.}\ \bibnamefont {Pan}},\ }\bibfield  {title} {\bibinfo
  {title} {Realization of two-dimensional spin-orbit coupling for bose-einstein
  condensates},\ }\href@noop {} {\bibfield  {journal} {\bibinfo  {journal}
  {Science}\ }\textbf {\bibinfo {volume} {354}},\ \bibinfo {pages} {83}
  (\bibinfo {year} {2016})}\BibitemShut {NoStop}%
\bibitem [{\citenamefont {Engelhardt}\ and\ \citenamefont
  {Brandes}(2015)}]{Engelhardt2015}%
  \BibitemOpen
  \bibfield  {author} {\bibinfo {author} {\bibfnamefont {G.}~\bibnamefont
  {Engelhardt}}\ and\ \bibinfo {author} {\bibfnamefont {T.}~\bibnamefont
  {Brandes}},\ }\bibfield  {title} {\bibinfo {title} {Topological bogoliubov
  excitations in inversion-symmetric systems of interacting bosons},\ }\href
  {https://doi.org/10.1103/PhysRevA.91.053621} {\bibfield  {journal} {\bibinfo
  {journal} {Phys. Rev. A}\ }\textbf {\bibinfo {volume} {91}},\ \bibinfo
  {pages} {053621} (\bibinfo {year} {2015})}\BibitemShut {NoStop}%
\bibitem [{\citenamefont {Furukawa}\ and\ \citenamefont
  {Ueda}(2015)}]{Furukawa2015}%
  \BibitemOpen
  \bibfield  {author} {\bibinfo {author} {\bibfnamefont {S.}~\bibnamefont
  {Furukawa}}\ and\ \bibinfo {author} {\bibfnamefont {M.}~\bibnamefont
  {Ueda}},\ }\bibfield  {title} {\bibinfo {title} {Excitation band topology and
  edge matter waves in bose{\textendash}einstein condensates in optical
  lattices},\ }\href {https://doi.org/10.1088/1367-2630/17/11/115014}
  {\bibfield  {journal} {\bibinfo  {journal} {New Journal of Physics}\ }\textbf
  {\bibinfo {volume} {17}},\ \bibinfo {pages} {115014} (\bibinfo {year}
  {2015})}\BibitemShut {NoStop}%
\bibitem [{\citenamefont {Aidelsburger}\ \emph {et~al.}(2015)\citenamefont
  {Aidelsburger}, \citenamefont {Lohse}, \citenamefont {Schweizer},
  \citenamefont {Atala}, \citenamefont {Barreiro}, \citenamefont
  {Nascimb{\`e}ne}, \citenamefont {Cooper}, \citenamefont {Bloch},\ and\
  \citenamefont {Goldman}}]{aidelsburger2015measuring}%
  \BibitemOpen
  \bibfield  {author} {\bibinfo {author} {\bibfnamefont {M.}~\bibnamefont
  {Aidelsburger}}, \bibinfo {author} {\bibfnamefont {M.}~\bibnamefont {Lohse}},
  \bibinfo {author} {\bibfnamefont {C.}~\bibnamefont {Schweizer}}, \bibinfo
  {author} {\bibfnamefont {M.}~\bibnamefont {Atala}}, \bibinfo {author}
  {\bibfnamefont {J.~T.}\ \bibnamefont {Barreiro}}, \bibinfo {author}
  {\bibfnamefont {S.}~\bibnamefont {Nascimb{\`e}ne}}, \bibinfo {author}
  {\bibfnamefont {N.}~\bibnamefont {Cooper}}, \bibinfo {author} {\bibfnamefont
  {I.}~\bibnamefont {Bloch}},\ and\ \bibinfo {author} {\bibfnamefont
  {N.}~\bibnamefont {Goldman}},\ }\bibfield  {title} {\bibinfo {title}
  {Measuring the chern number of hofstadter bands with ultracold bosonic
  atoms},\ }\href@noop {} {\bibfield  {journal} {\bibinfo  {journal} {Nature
  Physics}\ }\textbf {\bibinfo {volume} {11}},\ \bibinfo {pages} {162}
  (\bibinfo {year} {2015})}\BibitemShut {NoStop}%
\bibitem [{\citenamefont {Hadad}\ \emph {et~al.}(2018)\citenamefont {Hadad},
  \citenamefont {Soric}, \citenamefont {Khanikaev},\ and\ \citenamefont
  {Al{\`{u}}}}]{HadadNatElec2018}%
  \BibitemOpen
  \bibfield  {author} {\bibinfo {author} {\bibfnamefont {Y.}~\bibnamefont
  {Hadad}}, \bibinfo {author} {\bibfnamefont {J.~C.}\ \bibnamefont {Soric}},
  \bibinfo {author} {\bibfnamefont {A.~B.}\ \bibnamefont {Khanikaev}},\ and\
  \bibinfo {author} {\bibfnamefont {A.}~\bibnamefont {Al{\`{u}}}},\ }\bibfield
  {title} {\bibinfo {title} {{Self-induced topological protection in nonlinear
  circuit arrays}},\ }\href {https://doi.org/10.1038/s41928-018-0042-z}
  {\bibfield  {journal} {\bibinfo  {journal} {Nature Electronics}\ }\textbf
  {\bibinfo {volume} {1}},\ \bibinfo {pages} {178} (\bibinfo {year}
  {2018})}\BibitemShut {NoStop}%
\bibitem [{\citenamefont {Maczewsky}\ \emph {et~al.}(2020)\citenamefont
  {Maczewsky}, \citenamefont {Heinrich}, \citenamefont {Kremer}, \citenamefont
  {Ivanov}, \citenamefont {Ehrhardt}, \citenamefont {Martinez}, \citenamefont
  {Kartashov}, \citenamefont {Konotop}, \citenamefont {Torner}, \citenamefont
  {Bauer} \emph {et~al.}}]{maczewsky2020nonlinearity}%
  \BibitemOpen
  \bibfield  {author} {\bibinfo {author} {\bibfnamefont {L.~J.}\ \bibnamefont
  {Maczewsky}}, \bibinfo {author} {\bibfnamefont {M.}~\bibnamefont {Heinrich}},
  \bibinfo {author} {\bibfnamefont {M.}~\bibnamefont {Kremer}}, \bibinfo
  {author} {\bibfnamefont {S.~K.}\ \bibnamefont {Ivanov}}, \bibinfo {author}
  {\bibfnamefont {M.}~\bibnamefont {Ehrhardt}}, \bibinfo {author}
  {\bibfnamefont {F.}~\bibnamefont {Martinez}}, \bibinfo {author}
  {\bibfnamefont {Y.~V.}\ \bibnamefont {Kartashov}}, \bibinfo {author}
  {\bibfnamefont {V.~V.}\ \bibnamefont {Konotop}}, \bibinfo {author}
  {\bibfnamefont {L.}~\bibnamefont {Torner}}, \bibinfo {author} {\bibfnamefont
  {D.}~\bibnamefont {Bauer}}, \emph {et~al.},\ }\bibfield  {title} {\bibinfo
  {title} {Nonlinearity-induced photonic topological insulator},\ }\href@noop
  {} {\bibfield  {journal} {\bibinfo  {journal} {Science}\ }\textbf {\bibinfo
  {volume} {370}},\ \bibinfo {pages} {701} (\bibinfo {year}
  {2020})}\BibitemShut {NoStop}%
\bibitem [{\citenamefont {Lumer}\ \emph {et~al.}(2013)\citenamefont {Lumer},
  \citenamefont {Plotnik}, \citenamefont {Rechtsman},\ and\ \citenamefont
  {Segev}}]{Lumer2013}%
  \BibitemOpen
  \bibfield  {author} {\bibinfo {author} {\bibfnamefont {Y.}~\bibnamefont
  {Lumer}}, \bibinfo {author} {\bibfnamefont {Y.}~\bibnamefont {Plotnik}},
  \bibinfo {author} {\bibfnamefont {M.~C.}\ \bibnamefont {Rechtsman}},\ and\
  \bibinfo {author} {\bibfnamefont {M.}~\bibnamefont {Segev}},\ }\bibfield
  {title} {\bibinfo {title} {Self-localized states in photonic topological
  insulators},\ }\href {https://doi.org/10.1103/PhysRevLett.111.243905}
  {\bibfield  {journal} {\bibinfo  {journal} {Phys. Rev. Lett.}\ }\textbf
  {\bibinfo {volume} {111}},\ \bibinfo {pages} {243905} (\bibinfo {year}
  {2013})}\BibitemShut {NoStop}%
\bibitem [{\citenamefont {Solnyshkov}\ \emph {et~al.}(2017)\citenamefont
  {Solnyshkov}, \citenamefont {Bleu}, \citenamefont {Teklu},\ and\
  \citenamefont {Malpuech}}]{Solnyshkov2017}%
  \BibitemOpen
  \bibfield  {author} {\bibinfo {author} {\bibfnamefont {D.~D.}\ \bibnamefont
  {Solnyshkov}}, \bibinfo {author} {\bibfnamefont {O.}~\bibnamefont {Bleu}},
  \bibinfo {author} {\bibfnamefont {B.}~\bibnamefont {Teklu}},\ and\ \bibinfo
  {author} {\bibfnamefont {G.}~\bibnamefont {Malpuech}},\ }\bibfield  {title}
  {\bibinfo {title} {{Chirality of Topological Gap Solitons in Bosonic Dimer
  Chains}},\ }\href {https://doi.org/10.1103/PhysRevLett.118.023901} {\bibfield
   {journal} {\bibinfo  {journal} {Physical Review Letters}\ }\textbf {\bibinfo
  {volume} {118}},\ \bibinfo {pages} {023901} (\bibinfo {year} {2017})},\
  \Eprint {https://arxiv.org/abs/1607.01805} {arXiv:1607.01805} \BibitemShut
  {NoStop}%
\bibitem [{\citenamefont {Smirnova}\ \emph {et~al.}(2020)\citenamefont
  {Smirnova}, \citenamefont {Leykam}, \citenamefont {Chong},\ and\
  \citenamefont {Kivshar}}]{Smirnova2020}%
  \BibitemOpen
  \bibfield  {author} {\bibinfo {author} {\bibfnamefont {D.}~\bibnamefont
  {Smirnova}}, \bibinfo {author} {\bibfnamefont {D.}~\bibnamefont {Leykam}},
  \bibinfo {author} {\bibfnamefont {Y.}~\bibnamefont {Chong}},\ and\ \bibinfo
  {author} {\bibfnamefont {Y.}~\bibnamefont {Kivshar}},\ }\bibfield  {title}
  {\bibinfo {title} {Nonlinear topological photonics},\ }\href
  {https://doi.org/10.1063/1.5142397} {\bibfield  {journal} {\bibinfo
  {journal} {Applied Physics Reviews}\ }\textbf {\bibinfo {volume} {7}},\
  \bibinfo {pages} {021306} (\bibinfo {year} {2020})},\ \Eprint
  {https://arxiv.org/abs/https://doi.org/10.1063/1.5142397}
  {https://doi.org/10.1063/1.5142397} \BibitemShut {NoStop}%
\bibitem [{\citenamefont {Kirsch}\ \emph {et~al.}(2021)\citenamefont {Kirsch},
  \citenamefont {Zhang}, \citenamefont {Kremer}, \citenamefont {Maczewsky},
  \citenamefont {Ivanov}, \citenamefont {Kartashov}, \citenamefont {Torner},
  \citenamefont {Bauer}, \citenamefont {Szameit},\ and\ \citenamefont
  {Heinrich}}]{kirsch2021nonlinear}%
  \BibitemOpen
  \bibfield  {author} {\bibinfo {author} {\bibfnamefont {M.~S.}\ \bibnamefont
  {Kirsch}}, \bibinfo {author} {\bibfnamefont {Y.}~\bibnamefont {Zhang}},
  \bibinfo {author} {\bibfnamefont {M.}~\bibnamefont {Kremer}}, \bibinfo
  {author} {\bibfnamefont {L.~J.}\ \bibnamefont {Maczewsky}}, \bibinfo {author}
  {\bibfnamefont {S.~K.}\ \bibnamefont {Ivanov}}, \bibinfo {author}
  {\bibfnamefont {Y.~V.}\ \bibnamefont {Kartashov}}, \bibinfo {author}
  {\bibfnamefont {L.}~\bibnamefont {Torner}}, \bibinfo {author} {\bibfnamefont
  {D.}~\bibnamefont {Bauer}}, \bibinfo {author} {\bibfnamefont
  {A.}~\bibnamefont {Szameit}},\ and\ \bibinfo {author} {\bibfnamefont
  {M.}~\bibnamefont {Heinrich}},\ }\bibfield  {title} {\bibinfo {title}
  {Nonlinear second-order photonic topological insulators},\ }\href@noop {}
  {\bibfield  {journal} {\bibinfo  {journal} {Nat. Phys.}\ }\textbf {\bibinfo
  {volume} {17}},\ \bibinfo {pages} {995} (\bibinfo {year} {2021})}\BibitemShut
  {NoStop}%
\bibitem [{\citenamefont {Mukherjee}\ and\ \citenamefont
  {Rechtsman}(2020)}]{mukherjee2020observation}%
  \BibitemOpen
  \bibfield  {author} {\bibinfo {author} {\bibfnamefont {S.}~\bibnamefont
  {Mukherjee}}\ and\ \bibinfo {author} {\bibfnamefont {M.~C.}\ \bibnamefont
  {Rechtsman}},\ }\bibfield  {title} {\bibinfo {title} {Observation of floquet
  solitons in a topological bandgap},\ }\href@noop {} {\bibfield  {journal}
  {\bibinfo  {journal} {Science}\ }\textbf {\bibinfo {volume} {368}},\ \bibinfo
  {pages} {856} (\bibinfo {year} {2020})}\BibitemShut {NoStop}%
\bibitem [{\citenamefont {Guo}\ \emph {et~al.}(2020)\citenamefont {Guo},
  \citenamefont {Xia}, \citenamefont {Wang}, \citenamefont {Song},
  \citenamefont {Chen},\ and\ \citenamefont {Yang}}]{Guo2020}%
  \BibitemOpen
  \bibfield  {author} {\bibinfo {author} {\bibfnamefont {M.}~\bibnamefont
  {Guo}}, \bibinfo {author} {\bibfnamefont {S.}~\bibnamefont {Xia}}, \bibinfo
  {author} {\bibfnamefont {N.}~\bibnamefont {Wang}}, \bibinfo {author}
  {\bibfnamefont {D.}~\bibnamefont {Song}}, \bibinfo {author} {\bibfnamefont
  {Z.}~\bibnamefont {Chen}},\ and\ \bibinfo {author} {\bibfnamefont
  {J.}~\bibnamefont {Yang}},\ }\bibfield  {title} {\bibinfo {title} {Weakly
  nonlinear topological gap solitons in su--schrieffer--heeger photonic
  lattices},\ }\href {https://doi.org/10.1364/OL.411102} {\bibfield  {journal}
  {\bibinfo  {journal} {Opt. Lett.}\ }\textbf {\bibinfo {volume} {45}},\
  \bibinfo {pages} {6466} (\bibinfo {year} {2020})}\BibitemShut {NoStop}%
\bibitem [{\citenamefont {Pernet}\ \emph {et~al.}(2021)\citenamefont {Pernet},
  \citenamefont {St-Jean}, \citenamefont {Solnyshkov}, \citenamefont
  {Malpuech}, \citenamefont {Zambon}, \citenamefont {Real}, \citenamefont
  {Jamadi}, \citenamefont {Lemaître}, \citenamefont {Morassi}, \citenamefont
  {Gratiet}, \citenamefont {Baptiste}, \citenamefont {Harouri}, \citenamefont
  {Sagnes}, \citenamefont {Amo}, \citenamefont {Ravets},\ and\ \citenamefont
  {Bloch}}]{pernet2021topological}%
  \BibitemOpen
  \bibfield  {author} {\bibinfo {author} {\bibfnamefont {N.}~\bibnamefont
  {Pernet}}, \bibinfo {author} {\bibfnamefont {P.}~\bibnamefont {St-Jean}},
  \bibinfo {author} {\bibfnamefont {D.~D.}\ \bibnamefont {Solnyshkov}},
  \bibinfo {author} {\bibfnamefont {G.}~\bibnamefont {Malpuech}}, \bibinfo
  {author} {\bibfnamefont {N.~C.}\ \bibnamefont {Zambon}}, \bibinfo {author}
  {\bibfnamefont {B.}~\bibnamefont {Real}}, \bibinfo {author} {\bibfnamefont
  {O.}~\bibnamefont {Jamadi}}, \bibinfo {author} {\bibfnamefont
  {A.}~\bibnamefont {Lemaître}}, \bibinfo {author} {\bibfnamefont
  {M.}~\bibnamefont {Morassi}}, \bibinfo {author} {\bibfnamefont {L.~L.}\
  \bibnamefont {Gratiet}}, \bibinfo {author} {\bibfnamefont {T.}~\bibnamefont
  {Baptiste}}, \bibinfo {author} {\bibfnamefont {A.}~\bibnamefont {Harouri}},
  \bibinfo {author} {\bibfnamefont {I.}~\bibnamefont {Sagnes}}, \bibinfo
  {author} {\bibfnamefont {A.}~\bibnamefont {Amo}}, \bibinfo {author}
  {\bibfnamefont {S.}~\bibnamefont {Ravets}},\ and\ \bibinfo {author}
  {\bibfnamefont {J.}~\bibnamefont {Bloch}},\ }\href@noop {} {\bibinfo {title}
  {Topological gap solitons in a 1d non-hermitian lattice}} (\bibinfo {year}
  {2021}),\ \Eprint {https://arxiv.org/abs/2101.01038} {arXiv:2101.01038
  [cond-mat.mes-hall]} \BibitemShut {NoStop}%
\bibitem [{\citenamefont {Bleu}\ \emph {et~al.}(2018)\citenamefont {Bleu},
  \citenamefont {Malpuech},\ and\ \citenamefont {Solnyshkov}}]{Bleu2018nc}%
  \BibitemOpen
  \bibfield  {author} {\bibinfo {author} {\bibfnamefont {O.}~\bibnamefont
  {Bleu}}, \bibinfo {author} {\bibfnamefont {G.}~\bibnamefont {Malpuech}},\
  and\ \bibinfo {author} {\bibfnamefont {D.~D.}\ \bibnamefont {Solnyshkov}},\
  }\bibfield  {title} {\bibinfo {title} {Robust quantum valley hall effect for
  vortices in an interacting bosonic quantum fluid},\ }\href
  {https://www.nature.com/articles/s41467-018-06520-7} {\bibfield  {journal}
  {\bibinfo  {journal} {Nature Comm}\ }\textbf {\bibinfo {volume} {9}},\
  \bibinfo {pages} {3991} (\bibinfo {year} {2018})}\BibitemShut {NoStop}%
\bibitem [{\citenamefont {Thouless}(1998)}]{Thouless1998}%
  \BibitemOpen
  \bibfield  {author} {\bibinfo {author} {\bibfnamefont {D.~J.}\ \bibnamefont
  {Thouless}},\ }\href@noop {} {\emph {\bibinfo {title} {Topological Quantum
  Numbers in \\Nonrelativistic Physics}}}\ (\bibinfo  {publisher} {World
  Scientific Publishing Co, Singapore},\ \bibinfo {year} {1998})\BibitemShut
  {NoStop}%
\bibitem [{\citenamefont {Kibble}(1976)}]{Kibble1976}%
  \BibitemOpen
  \bibfield  {author} {\bibinfo {author} {\bibfnamefont {T.}~\bibnamefont
  {Kibble}},\ }\href@noop {} {\bibfield  {journal} {\bibinfo  {journal} {J.
  Phys. A.:Math. Gen.}\ }\textbf {\bibinfo {volume} {9}},\ \bibinfo {pages}
  {1387} (\bibinfo {year} {1976})}\BibitemShut {NoStop}%
\bibitem [{\citenamefont {Zurek}(1985)}]{Zurek1985}%
  \BibitemOpen
  \bibfield  {author} {\bibinfo {author} {\bibfnamefont {W.}~\bibnamefont
  {Zurek}},\ }\href@noop {} {\bibfield  {journal} {\bibinfo  {journal}
  {Nature}\ }\textbf {\bibinfo {volume} {317}},\ \bibinfo {pages} {505}
  (\bibinfo {year} {1985})}\BibitemShut {NoStop}%
\bibitem [{\citenamefont {Zurek}(1996)}]{Zurek1996}%
  \BibitemOpen
  \bibfield  {author} {\bibinfo {author} {\bibfnamefont {W.}~\bibnamefont
  {Zurek}},\ }\href@noop {} {\bibfield  {journal} {\bibinfo  {journal} {Physics
  Reports}\ }\textbf {\bibinfo {volume} {276}},\ \bibinfo {pages} {177}
  (\bibinfo {year} {1996})}\BibitemShut {NoStop}%
\bibitem [{\citenamefont {Yao}\ \emph {et~al.}(2022)\citenamefont {Yao},
  \citenamefont {Zhang},\ and\ \citenamefont {Chin}}]{yao2022domain}%
  \BibitemOpen
  \bibfield  {author} {\bibinfo {author} {\bibfnamefont {K.-X.}\ \bibnamefont
  {Yao}}, \bibinfo {author} {\bibfnamefont {Z.}~\bibnamefont {Zhang}},\ and\
  \bibinfo {author} {\bibfnamefont {C.}~\bibnamefont {Chin}},\ }\bibfield
  {title} {\bibinfo {title} {Domain-wall dynamics in bose--einstein condensates
  with synthetic gauge fields},\ }\href@noop {} {\bibfield  {journal} {\bibinfo
   {journal} {Nature}\ }\textbf {\bibinfo {volume} {602}},\ \bibinfo {pages}
  {68} (\bibinfo {year} {2022})}\BibitemShut {NoStop}%
\bibitem [{\citenamefont {Jackiw}\ and\ \citenamefont
  {Rebbi}(1976)}]{Jackiw1976}%
  \BibitemOpen
  \bibfield  {author} {\bibinfo {author} {\bibfnamefont {R.}~\bibnamefont
  {Jackiw}}\ and\ \bibinfo {author} {\bibfnamefont {C.}~\bibnamefont {Rebbi}},\
  }\bibfield  {title} {\bibinfo {title} {Solitons with fermion number
  \textonehalf{}},\ }\href {https://doi.org/10.1103/PhysRevD.13.3398}
  {\bibfield  {journal} {\bibinfo  {journal} {Phys. Rev. D}\ }\textbf {\bibinfo
  {volume} {13}},\ \bibinfo {pages} {3398} (\bibinfo {year}
  {1976})}\BibitemShut {NoStop}%
\bibitem [{\citenamefont {Jacqmin}\ \emph {et~al.}(2014)\citenamefont
  {Jacqmin}, \citenamefont {Carusotto}, \citenamefont {Sagnes}, \citenamefont
  {Abbarchi}, \citenamefont {Solnyshkov}, \citenamefont {Malpuech},
  \citenamefont {Galopin}, \citenamefont {Lema\^{\i}tre}, \citenamefont
  {Bloch},\ and\ \citenamefont {Amo}}]{Jacqmin2014}%
  \BibitemOpen
  \bibfield  {author} {\bibinfo {author} {\bibfnamefont {T.}~\bibnamefont
  {Jacqmin}}, \bibinfo {author} {\bibfnamefont {I.}~\bibnamefont {Carusotto}},
  \bibinfo {author} {\bibfnamefont {I.}~\bibnamefont {Sagnes}}, \bibinfo
  {author} {\bibfnamefont {M.}~\bibnamefont {Abbarchi}}, \bibinfo {author}
  {\bibfnamefont {D.~D.}\ \bibnamefont {Solnyshkov}}, \bibinfo {author}
  {\bibfnamefont {G.}~\bibnamefont {Malpuech}}, \bibinfo {author}
  {\bibfnamefont {E.}~\bibnamefont {Galopin}}, \bibinfo {author} {\bibfnamefont
  {A.}~\bibnamefont {Lema\^{\i}tre}}, \bibinfo {author} {\bibfnamefont
  {J.}~\bibnamefont {Bloch}},\ and\ \bibinfo {author} {\bibfnamefont
  {A.}~\bibnamefont {Amo}},\ }\bibfield  {title} {\bibinfo {title} {Direct
  observation of {D}irac cones and a flatband in a honeycomb lattice for
  polaritons},\ }\href {https://doi.org/10.1103/PhysRevLett.112.116402}
  {\bibfield  {journal} {\bibinfo  {journal} {Phys. Rev. Lett.}\ }\textbf
  {\bibinfo {volume} {112}},\ \bibinfo {pages} {116402} (\bibinfo {year}
  {2014})}\BibitemShut {NoStop}%
\bibitem [{\citenamefont {Klembt}\ \emph {et~al.}(2017)\citenamefont {Klembt},
  \citenamefont {Harder}, \citenamefont {Egorov}, \citenamefont {Winkler},
  \citenamefont {Suchomel}, \citenamefont {Beierlein}, \citenamefont
  {Emmerling}, \citenamefont {Schneider},\ and\ \citenamefont
  {H{\"o}fling}}]{klembt2017polariton}%
  \BibitemOpen
  \bibfield  {author} {\bibinfo {author} {\bibfnamefont {S.}~\bibnamefont
  {Klembt}}, \bibinfo {author} {\bibfnamefont {T.~H.}\ \bibnamefont {Harder}},
  \bibinfo {author} {\bibfnamefont {O.~A.}\ \bibnamefont {Egorov}}, \bibinfo
  {author} {\bibfnamefont {K.}~\bibnamefont {Winkler}}, \bibinfo {author}
  {\bibfnamefont {H.}~\bibnamefont {Suchomel}}, \bibinfo {author}
  {\bibfnamefont {J.}~\bibnamefont {Beierlein}}, \bibinfo {author}
  {\bibfnamefont {M.}~\bibnamefont {Emmerling}}, \bibinfo {author}
  {\bibfnamefont {C.}~\bibnamefont {Schneider}},\ and\ \bibinfo {author}
  {\bibfnamefont {S.}~\bibnamefont {H{\"o}fling}},\ }\bibfield  {title}
  {\bibinfo {title} {Polariton condensation in s-and p-flatbands in a
  two-dimensional lieb lattice},\ }\href@noop {} {\bibfield  {journal}
  {\bibinfo  {journal} {Applied Physics Letters}\ }\textbf {\bibinfo {volume}
  {111}},\ \bibinfo {pages} {231102} (\bibinfo {year} {2017})}\BibitemShut
  {NoStop}%
\bibitem [{\citenamefont {Whittaker}\ \emph {et~al.}(2018)\citenamefont
  {Whittaker}, \citenamefont {Cancellieri}, \citenamefont {Walker},
  \citenamefont {Gulevich}, \citenamefont {Schomerus}, \citenamefont
  {Vaitiekus}, \citenamefont {Royall}, \citenamefont {Whittaker}, \citenamefont
  {Clarke}, \citenamefont {Iorsh}, \citenamefont {Shelykh}, \citenamefont
  {Skolnick},\ and\ \citenamefont {Krizhanovskii}}]{Whittaker2018}%
  \BibitemOpen
  \bibfield  {author} {\bibinfo {author} {\bibfnamefont {C.~E.}\ \bibnamefont
  {Whittaker}}, \bibinfo {author} {\bibfnamefont {E.}~\bibnamefont
  {Cancellieri}}, \bibinfo {author} {\bibfnamefont {P.~M.}\ \bibnamefont
  {Walker}}, \bibinfo {author} {\bibfnamefont {D.~R.}\ \bibnamefont
  {Gulevich}}, \bibinfo {author} {\bibfnamefont {H.}~\bibnamefont {Schomerus}},
  \bibinfo {author} {\bibfnamefont {D.}~\bibnamefont {Vaitiekus}}, \bibinfo
  {author} {\bibfnamefont {B.}~\bibnamefont {Royall}}, \bibinfo {author}
  {\bibfnamefont {D.~M.}\ \bibnamefont {Whittaker}}, \bibinfo {author}
  {\bibfnamefont {E.}~\bibnamefont {Clarke}}, \bibinfo {author} {\bibfnamefont
  {I.~V.}\ \bibnamefont {Iorsh}}, \bibinfo {author} {\bibfnamefont {I.~A.}\
  \bibnamefont {Shelykh}}, \bibinfo {author} {\bibfnamefont {M.~S.}\
  \bibnamefont {Skolnick}},\ and\ \bibinfo {author} {\bibfnamefont {D.~N.}\
  \bibnamefont {Krizhanovskii}},\ }\bibfield  {title} {\bibinfo {title}
  {Exciton polaritons in a two-dimensional {L}ieb lattice with spin-orbit
  coupling},\ }\href {https://doi.org/10.1103/PhysRevLett.120.097401}
  {\bibfield  {journal} {\bibinfo  {journal} {Phys. Rev. Lett.}\ }\textbf
  {\bibinfo {volume} {120}},\ \bibinfo {pages} {097401} (\bibinfo {year}
  {2018})}\BibitemShut {NoStop}%
\bibitem [{\citenamefont {Real}\ \emph {et~al.}(2020)\citenamefont {Real},
  \citenamefont {Jamadi}, \citenamefont {Mili\ifmmode \acute{c}\else
  \'{c}\fi{}evi\ifmmode~\acute{c}\else \'{c}\fi{}}, \citenamefont {Pernet},
  \citenamefont {St-Jean}, \citenamefont {Ozawa}, \citenamefont {Montambaux},
  \citenamefont {Sagnes}, \citenamefont {Lema\^{\i}tre}, \citenamefont
  {Le~Gratiet}, \citenamefont {Harouri}, \citenamefont {Ravets}, \citenamefont
  {Bloch},\ and\ \citenamefont {Amo}}]{Real2020}%
  \BibitemOpen
  \bibfield  {author} {\bibinfo {author} {\bibfnamefont {B.}~\bibnamefont
  {Real}}, \bibinfo {author} {\bibfnamefont {O.}~\bibnamefont {Jamadi}},
  \bibinfo {author} {\bibfnamefont {M.}~\bibnamefont {Mili\ifmmode
  \acute{c}\else \'{c}\fi{}evi\ifmmode~\acute{c}\else \'{c}\fi{}}}, \bibinfo
  {author} {\bibfnamefont {N.}~\bibnamefont {Pernet}}, \bibinfo {author}
  {\bibfnamefont {P.}~\bibnamefont {St-Jean}}, \bibinfo {author} {\bibfnamefont
  {T.}~\bibnamefont {Ozawa}}, \bibinfo {author} {\bibfnamefont
  {G.}~\bibnamefont {Montambaux}}, \bibinfo {author} {\bibfnamefont
  {I.}~\bibnamefont {Sagnes}}, \bibinfo {author} {\bibfnamefont
  {A.}~\bibnamefont {Lema\^{\i}tre}}, \bibinfo {author} {\bibfnamefont
  {L.}~\bibnamefont {Le~Gratiet}}, \bibinfo {author} {\bibfnamefont
  {A.}~\bibnamefont {Harouri}}, \bibinfo {author} {\bibfnamefont
  {S.}~\bibnamefont {Ravets}}, \bibinfo {author} {\bibfnamefont
  {J.}~\bibnamefont {Bloch}},\ and\ \bibinfo {author} {\bibfnamefont
  {A.}~\bibnamefont {Amo}},\ }\bibfield  {title} {\bibinfo {title} {Semi-dirac
  transport and anisotropic localization in polariton honeycomb lattices},\
  }\href {https://doi.org/10.1103/PhysRevLett.125.186601} {\bibfield  {journal}
  {\bibinfo  {journal} {Phys. Rev. Lett.}\ }\textbf {\bibinfo {volume} {125}},\
  \bibinfo {pages} {186601} (\bibinfo {year} {2020})}\BibitemShut {NoStop}%
\bibitem [{\citenamefont {Mili\ifmmode \acute{c}\else
  \'{c}\fi{}evi\ifmmode~\acute{c}\else \'{c}\fi{}}\ \emph
  {et~al.}(2019)\citenamefont {Mili\ifmmode \acute{c}\else
  \'{c}\fi{}evi\ifmmode~\acute{c}\else \'{c}\fi{}}, \citenamefont {Montambaux},
  \citenamefont {Ozawa}, \citenamefont {Jamadi}, \citenamefont {Real},
  \citenamefont {Sagnes}, \citenamefont {Lema\^{\i}tre}, \citenamefont
  {Le~Gratiet}, \citenamefont {Harouri}, \citenamefont {Bloch},\ and\
  \citenamefont {Amo}}]{Milicevic2019}%
  \BibitemOpen
  \bibfield  {author} {\bibinfo {author} {\bibfnamefont {M.}~\bibnamefont
  {Mili\ifmmode \acute{c}\else \'{c}\fi{}evi\ifmmode~\acute{c}\else
  \'{c}\fi{}}}, \bibinfo {author} {\bibfnamefont {G.}~\bibnamefont
  {Montambaux}}, \bibinfo {author} {\bibfnamefont {T.}~\bibnamefont {Ozawa}},
  \bibinfo {author} {\bibfnamefont {O.}~\bibnamefont {Jamadi}}, \bibinfo
  {author} {\bibfnamefont {B.}~\bibnamefont {Real}}, \bibinfo {author}
  {\bibfnamefont {I.}~\bibnamefont {Sagnes}}, \bibinfo {author} {\bibfnamefont
  {A.}~\bibnamefont {Lema\^{\i}tre}}, \bibinfo {author} {\bibfnamefont
  {L.}~\bibnamefont {Le~Gratiet}}, \bibinfo {author} {\bibfnamefont
  {A.}~\bibnamefont {Harouri}}, \bibinfo {author} {\bibfnamefont
  {J.}~\bibnamefont {Bloch}},\ and\ \bibinfo {author} {\bibfnamefont
  {A.}~\bibnamefont {Amo}},\ }\bibfield  {title} {\bibinfo {title} {Type-iii
  and tilted dirac cones emerging from flat bands in photonic orbital
  graphene},\ }\href {https://doi.org/10.1103/PhysRevX.9.031010} {\bibfield
  {journal} {\bibinfo  {journal} {Phys. Rev. X}\ }\textbf {\bibinfo {volume}
  {9}},\ \bibinfo {pages} {031010} (\bibinfo {year} {2019})}\BibitemShut
  {NoStop}%
\bibitem [{\citenamefont {Klembt}\ \emph {et~al.}(2018)\citenamefont {Klembt},
  \citenamefont {Harder}, \citenamefont {Egorov}, \citenamefont {Winkler},
  \citenamefont {Ge}, \citenamefont {Bandres}, \citenamefont {Emmerling},
  \citenamefont {Worschech}, \citenamefont {Liew}, \citenamefont {Segev} \emph
  {et~al.}}]{klembt2018exciton}%
  \BibitemOpen
  \bibfield  {author} {\bibinfo {author} {\bibfnamefont {S.}~\bibnamefont
  {Klembt}}, \bibinfo {author} {\bibfnamefont {T.}~\bibnamefont {Harder}},
  \bibinfo {author} {\bibfnamefont {O.}~\bibnamefont {Egorov}}, \bibinfo
  {author} {\bibfnamefont {K.}~\bibnamefont {Winkler}}, \bibinfo {author}
  {\bibfnamefont {R.}~\bibnamefont {Ge}}, \bibinfo {author} {\bibfnamefont
  {M.}~\bibnamefont {Bandres}}, \bibinfo {author} {\bibfnamefont
  {M.}~\bibnamefont {Emmerling}}, \bibinfo {author} {\bibfnamefont
  {L.}~\bibnamefont {Worschech}}, \bibinfo {author} {\bibfnamefont
  {T.}~\bibnamefont {Liew}}, \bibinfo {author} {\bibfnamefont {M.}~\bibnamefont
  {Segev}}, \emph {et~al.},\ }\bibfield  {title} {\bibinfo {title}
  {Exciton-polariton topological insulator},\ }\href@noop {} {\bibfield
  {journal} {\bibinfo  {journal} {Nature}\ }\textbf {\bibinfo {volume} {562}},\
  \bibinfo {pages} {552} (\bibinfo {year} {2018})}\BibitemShut {NoStop}%
\bibitem [{\citenamefont {Suchomel}\ \emph {et~al.}(2018)\citenamefont
  {Suchomel}, \citenamefont {Klembt}, \citenamefont {Harder}, \citenamefont
  {Klaas}, \citenamefont {Egorov}, \citenamefont {Winkler}, \citenamefont
  {Emmerling}, \citenamefont {Thomale}, \citenamefont {H{\"{o}}fling},\ and\
  \citenamefont {Schneider}}]{Suchomel2018}%
  \BibitemOpen
  \bibfield  {author} {\bibinfo {author} {\bibfnamefont {H.}~\bibnamefont
  {Suchomel}}, \bibinfo {author} {\bibfnamefont {S.}~\bibnamefont {Klembt}},
  \bibinfo {author} {\bibfnamefont {T.~H.}\ \bibnamefont {Harder}}, \bibinfo
  {author} {\bibfnamefont {M.}~\bibnamefont {Klaas}}, \bibinfo {author}
  {\bibfnamefont {O.~A.}\ \bibnamefont {Egorov}}, \bibinfo {author}
  {\bibfnamefont {K.}~\bibnamefont {Winkler}}, \bibinfo {author} {\bibfnamefont
  {M.}~\bibnamefont {Emmerling}}, \bibinfo {author} {\bibfnamefont
  {R.}~\bibnamefont {Thomale}}, \bibinfo {author} {\bibfnamefont
  {S.}~\bibnamefont {H{\"{o}}fling}},\ and\ \bibinfo {author} {\bibfnamefont
  {C.}~\bibnamefont {Schneider}},\ }\bibfield  {title} {\bibinfo {title}
  {{Platform for Electrically Pumped Polariton Simulators and Topological
  Lasers}},\ }\href {https://doi.org/10.1103/PhysRevLett.121.257402} {\bibfield
   {journal} {\bibinfo  {journal} {Physical Review Letters}\ }\textbf {\bibinfo
  {volume} {121}},\ \bibinfo {pages} {257402} (\bibinfo {year}
  {2018})}\BibitemShut {NoStop}%
\bibitem [{\citenamefont {Lledo}\ \emph {et~al.}(2021)\citenamefont {Lledo},
  \citenamefont {Carusotto},\ and\ \citenamefont {Szymanska}}]{Lledo2021}%
  \BibitemOpen
  \bibfield  {author} {\bibinfo {author} {\bibfnamefont {C.}~\bibnamefont
  {Lledo}}, \bibinfo {author} {\bibfnamefont {I.}~\bibnamefont {Carusotto}},\
  and\ \bibinfo {author} {\bibfnamefont {M.~H.}\ \bibnamefont {Szymanska}},\
  }\bibfield  {title} {\bibinfo {title} {Polariton condensation into vortex
  states in the synthetic magnetic field of a strained honeycomb lattice},\
  }\href@noop {} {\bibfield  {journal} {\bibinfo  {journal}
  {scipost:202104.00031}\ } (\bibinfo {year} {2021})}\BibitemShut {NoStop}%
\bibitem [{\citenamefont {LP}(1959)}]{Pitaevskii58}%
  \BibitemOpen
  \bibfield  {author} {\bibinfo {author} {\bibfnamefont {P.}~\bibnamefont
  {LP}},\ }\bibfield  {title} {\bibinfo {title} {Phenomenological theory of
  superfluidity near the lambda point},\ }\href@noop {} {\bibfield  {journal}
  {\bibinfo  {journal} {Sov. Phys. JETP}\ }\textbf {\bibinfo {volume} {35}},\
  \bibinfo {pages} {282} (\bibinfo {year} {1959})}\BibitemShut {NoStop}%
\bibitem [{\citenamefont {Solnyshkov}\ \emph {et~al.}(2021)\citenamefont
  {Solnyshkov}, \citenamefont {Bessonart}, \citenamefont {Nalitov},\ and\
  \citenamefont {Malpuech}}]{GKZM}%
  \BibitemOpen
  \bibfield  {author} {\bibinfo {author} {\bibfnamefont {D.~D.}\ \bibnamefont
  {Solnyshkov}}, \bibinfo {author} {\bibfnamefont {L.}~\bibnamefont
  {Bessonart}}, \bibinfo {author} {\bibfnamefont {A.}~\bibnamefont {Nalitov}},\
  and\ \bibinfo {author} {\bibfnamefont {G.}~\bibnamefont {Malpuech}},\
  }\bibfield  {title} {\bibinfo {title} {Kibble-zurek mechanism in polariton
  graphene},\ }\href {https://doi.org/10.1103/PhysRevB.104.035423} {\bibfield
  {journal} {\bibinfo  {journal} {Phys. Rev. B}\ }\textbf {\bibinfo {volume}
  {104}},\ \bibinfo {pages} {035423} (\bibinfo {year} {2021})}\BibitemShut
  {NoStop}%
\bibitem [{\citenamefont {Kasprzak}\ \emph {et~al.}(2008)\citenamefont
  {Kasprzak}, \citenamefont {Solnyshkov}, \citenamefont {Andr\'e},
  \citenamefont {Dang},\ and\ \citenamefont {Malpuech}}]{Kasprzak2008}%
  \BibitemOpen
  \bibfield  {author} {\bibinfo {author} {\bibfnamefont {J.}~\bibnamefont
  {Kasprzak}}, \bibinfo {author} {\bibfnamefont {D.~D.}\ \bibnamefont
  {Solnyshkov}}, \bibinfo {author} {\bibfnamefont {R.}~\bibnamefont {Andr\'e}},
  \bibinfo {author} {\bibfnamefont {L.~S.}\ \bibnamefont {Dang}},\ and\
  \bibinfo {author} {\bibfnamefont {G.}~\bibnamefont {Malpuech}},\ }\bibfield
  {title} {\bibinfo {title} {Formation of an exciton polariton condensate:
  Thermodynamic versus kinetic regimes},\ }\href
  {https://doi.org/10.1103/PhysRevLett.101.146404} {\bibfield  {journal}
  {\bibinfo  {journal} {Phys. Rev. Lett.}\ }\textbf {\bibinfo {volume} {101}},\
  \bibinfo {pages} {146404} (\bibinfo {year} {2008})}\BibitemShut {NoStop}%
\bibitem [{\citenamefont {Levrat}\ \emph {et~al.}(2010)\citenamefont {Levrat},
  \citenamefont {Butt\'e}, \citenamefont {Feltin}, \citenamefont {Carlin},
  \citenamefont {Grandjean}, \citenamefont {Solnyshkov},\ and\ \citenamefont
  {Malpuech}}]{Levrat2010}%
  \BibitemOpen
  \bibfield  {author} {\bibinfo {author} {\bibfnamefont {J.}~\bibnamefont
  {Levrat}}, \bibinfo {author} {\bibfnamefont {R.}~\bibnamefont {Butt\'e}},
  \bibinfo {author} {\bibfnamefont {E.}~\bibnamefont {Feltin}}, \bibinfo
  {author} {\bibfnamefont {J.-F. m.~c.}\ \bibnamefont {Carlin}}, \bibinfo
  {author} {\bibfnamefont {N.}~\bibnamefont {Grandjean}}, \bibinfo {author}
  {\bibfnamefont {D.}~\bibnamefont {Solnyshkov}},\ and\ \bibinfo {author}
  {\bibfnamefont {G.}~\bibnamefont {Malpuech}},\ }\bibfield  {title} {\bibinfo
  {title} {Condensation phase diagram of cavity polaritons in gan-based
  microcavities: Experiment and theory},\ }\href
  {https://doi.org/10.1103/PhysRevB.81.125305} {\bibfield  {journal} {\bibinfo
  {journal} {Phys. Rev. B}\ }\textbf {\bibinfo {volume} {81}},\ \bibinfo
  {pages} {125305} (\bibinfo {year} {2010})}\BibitemShut {NoStop}%
\bibitem [{\citenamefont {Li}\ \emph {et~al.}(2013)\citenamefont {Li},
  \citenamefont {Orosz}, \citenamefont {Kamoun}, \citenamefont {Bouchoule},
  \citenamefont {Brimont}, \citenamefont {Disseix}, \citenamefont {Guillet},
  \citenamefont {Lafosse}, \citenamefont {Leroux}, \citenamefont {Leymarie},
  \citenamefont {Mexis}, \citenamefont {Mihailovic}, \citenamefont
  {Patriarche}, \citenamefont {R\'everet}, \citenamefont {Solnyshkov},
  \citenamefont {Zuniga-Perez},\ and\ \citenamefont {Malpuech}}]{Feng2013}%
  \BibitemOpen
  \bibfield  {author} {\bibinfo {author} {\bibfnamefont {F.}~\bibnamefont
  {Li}}, \bibinfo {author} {\bibfnamefont {L.}~\bibnamefont {Orosz}}, \bibinfo
  {author} {\bibfnamefont {O.}~\bibnamefont {Kamoun}}, \bibinfo {author}
  {\bibfnamefont {S.}~\bibnamefont {Bouchoule}}, \bibinfo {author}
  {\bibfnamefont {C.}~\bibnamefont {Brimont}}, \bibinfo {author} {\bibfnamefont
  {P.}~\bibnamefont {Disseix}}, \bibinfo {author} {\bibfnamefont
  {T.}~\bibnamefont {Guillet}}, \bibinfo {author} {\bibfnamefont
  {X.}~\bibnamefont {Lafosse}}, \bibinfo {author} {\bibfnamefont
  {M.}~\bibnamefont {Leroux}}, \bibinfo {author} {\bibfnamefont
  {J.}~\bibnamefont {Leymarie}}, \bibinfo {author} {\bibfnamefont
  {M.}~\bibnamefont {Mexis}}, \bibinfo {author} {\bibfnamefont
  {M.}~\bibnamefont {Mihailovic}}, \bibinfo {author} {\bibfnamefont
  {G.}~\bibnamefont {Patriarche}}, \bibinfo {author} {\bibfnamefont
  {F.}~\bibnamefont {R\'everet}}, \bibinfo {author} {\bibfnamefont
  {D.}~\bibnamefont {Solnyshkov}}, \bibinfo {author} {\bibfnamefont
  {J.}~\bibnamefont {Zuniga-Perez}},\ and\ \bibinfo {author} {\bibfnamefont
  {G.}~\bibnamefont {Malpuech}},\ }\bibfield  {title} {\bibinfo {title} {From
  excitonic to photonic polariton condensate in a zno-based microcavity},\
  }\href {https://doi.org/10.1103/PhysRevLett.110.196406} {\bibfield  {journal}
  {\bibinfo  {journal} {Phys. Rev. Lett.}\ }\textbf {\bibinfo {volume} {110}},\
  \bibinfo {pages} {196406} (\bibinfo {year} {2013})}\BibitemShut {NoStop}%
\bibitem [{\citenamefont {Wertz}\ \emph {et~al.}(2009)\citenamefont {Wertz},
  \citenamefont {Ferrier}, \citenamefont {Solnyshkov}, \citenamefont
  {Senellart}, \citenamefont {Bajoni}, \citenamefont {Miard}, \citenamefont
  {Lemaître}, \citenamefont {Malpuech},\ and\ \citenamefont
  {Bloch}}]{Wertz2009}%
  \BibitemOpen
  \bibfield  {author} {\bibinfo {author} {\bibfnamefont {E.}~\bibnamefont
  {Wertz}}, \bibinfo {author} {\bibfnamefont {L.}~\bibnamefont {Ferrier}},
  \bibinfo {author} {\bibfnamefont {D.~D.}\ \bibnamefont {Solnyshkov}},
  \bibinfo {author} {\bibfnamefont {P.}~\bibnamefont {Senellart}}, \bibinfo
  {author} {\bibfnamefont {D.}~\bibnamefont {Bajoni}}, \bibinfo {author}
  {\bibfnamefont {A.}~\bibnamefont {Miard}}, \bibinfo {author} {\bibfnamefont
  {A.}~\bibnamefont {Lemaître}}, \bibinfo {author} {\bibfnamefont
  {G.}~\bibnamefont {Malpuech}},\ and\ \bibinfo {author} {\bibfnamefont
  {J.}~\bibnamefont {Bloch}},\ }\bibfield  {title} {\bibinfo {title}
  {Spontaneous formation of a polariton condensate in a planar gaas
  microcavity},\ }\href {https://doi.org/10.1063/1.3192408} {\bibfield
  {journal} {\bibinfo  {journal} {Applied Physics Letters}\ }\textbf {\bibinfo
  {volume} {95}},\ \bibinfo {pages} {051108} (\bibinfo {year} {2009})},\
  \Eprint {https://arxiv.org/abs/https://doi.org/10.1063/1.3192408}
  {https://doi.org/10.1063/1.3192408} \BibitemShut {NoStop}%
\bibitem [{\citenamefont {Jamadi}\ \emph {et~al.}(2019)\citenamefont {Jamadi},
  \citenamefont {R\'everet}, \citenamefont {Solnyshkov}, \citenamefont
  {Disseix}, \citenamefont {Leymarie}, \citenamefont {Mallet-Dida},
  \citenamefont {Brimont}, \citenamefont {Guillet}, \citenamefont {Lafosse},
  \citenamefont {Bouchoule}, \citenamefont {Semond}, \citenamefont {Leroux},
  \citenamefont {Zuniga-Perez},\ and\ \citenamefont {Malpuech}}]{Jamadi2019}%
  \BibitemOpen
  \bibfield  {author} {\bibinfo {author} {\bibfnamefont {O.}~\bibnamefont
  {Jamadi}}, \bibinfo {author} {\bibfnamefont {F.}~\bibnamefont {R\'everet}},
  \bibinfo {author} {\bibfnamefont {D.}~\bibnamefont {Solnyshkov}}, \bibinfo
  {author} {\bibfnamefont {P.}~\bibnamefont {Disseix}}, \bibinfo {author}
  {\bibfnamefont {J.}~\bibnamefont {Leymarie}}, \bibinfo {author}
  {\bibfnamefont {L.}~\bibnamefont {Mallet-Dida}}, \bibinfo {author}
  {\bibfnamefont {C.}~\bibnamefont {Brimont}}, \bibinfo {author} {\bibfnamefont
  {T.}~\bibnamefont {Guillet}}, \bibinfo {author} {\bibfnamefont
  {X.}~\bibnamefont {Lafosse}}, \bibinfo {author} {\bibfnamefont
  {S.}~\bibnamefont {Bouchoule}}, \bibinfo {author} {\bibfnamefont
  {F.}~\bibnamefont {Semond}}, \bibinfo {author} {\bibfnamefont
  {M.}~\bibnamefont {Leroux}}, \bibinfo {author} {\bibfnamefont
  {J.}~\bibnamefont {Zuniga-Perez}},\ and\ \bibinfo {author} {\bibfnamefont
  {G.}~\bibnamefont {Malpuech}},\ }\bibfield  {title} {\bibinfo {title}
  {Competition between horizontal and vertical polariton lasing in planar
  microcavities},\ }\href {https://doi.org/10.1103/PhysRevB.99.085304}
  {\bibfield  {journal} {\bibinfo  {journal} {Phys. Rev. B}\ }\textbf {\bibinfo
  {volume} {99}},\ \bibinfo {pages} {085304} (\bibinfo {year}
  {2019})}\BibitemShut {NoStop}%
\bibitem [{\citenamefont {Milićević}\ \emph {et~al.}(2018)\citenamefont
  {Milićević}, \citenamefont {Bleu}, \citenamefont {Solnyshkov},
  \citenamefont {Sagnes}, \citenamefont {Lemaître}, \citenamefont {Gratiet},
  \citenamefont {Harouri}, \citenamefont {Bloch}, \citenamefont {Malpuech},\
  and\ \citenamefont {Amo}}]{Milic2018}%
  \BibitemOpen
  \bibfield  {author} {\bibinfo {author} {\bibfnamefont {M.}~\bibnamefont
  {Milićević}}, \bibinfo {author} {\bibfnamefont {O.}~\bibnamefont {Bleu}},
  \bibinfo {author} {\bibfnamefont {D.~D.}\ \bibnamefont {Solnyshkov}},
  \bibinfo {author} {\bibfnamefont {I.}~\bibnamefont {Sagnes}}, \bibinfo
  {author} {\bibfnamefont {A.}~\bibnamefont {Lemaître}}, \bibinfo {author}
  {\bibfnamefont {L.~L.}\ \bibnamefont {Gratiet}}, \bibinfo {author}
  {\bibfnamefont {A.}~\bibnamefont {Harouri}}, \bibinfo {author} {\bibfnamefont
  {J.}~\bibnamefont {Bloch}}, \bibinfo {author} {\bibfnamefont
  {G.}~\bibnamefont {Malpuech}},\ and\ \bibinfo {author} {\bibfnamefont
  {A.}~\bibnamefont {Amo}},\ }\bibfield  {title} {\bibinfo {title} {{Lasing in
  optically induced gap states in photonic graphene}},\ }\href
  {https://doi.org/10.21468/SciPostPhys.5.6.064} {\bibfield  {journal}
  {\bibinfo  {journal} {SciPost Phys.}\ }\textbf {\bibinfo {volume} {5}},\
  \bibinfo {pages} {64} (\bibinfo {year} {2018})}\BibitemShut {NoStop}%
\bibitem [{\citenamefont {Aleiner}\ \emph {et~al.}(2012)\citenamefont
  {Aleiner}, \citenamefont {Altshuler},\ and\ \citenamefont
  {Rubo}}]{Aleiner2012}%
  \BibitemOpen
  \bibfield  {author} {\bibinfo {author} {\bibfnamefont {I.~L.}\ \bibnamefont
  {Aleiner}}, \bibinfo {author} {\bibfnamefont {B.~L.}\ \bibnamefont
  {Altshuler}},\ and\ \bibinfo {author} {\bibfnamefont {Y.~G.}\ \bibnamefont
  {Rubo}},\ }\bibfield  {title} {\bibinfo {title} {Radiative coupling and weak
  lasing of exciton-polariton condensates},\ }\href
  {https://doi.org/10.1103/PhysRevB.85.121301} {\bibfield  {journal} {\bibinfo
  {journal} {Phys. Rev. B}\ }\textbf {\bibinfo {volume} {85}},\ \bibinfo
  {pages} {121301} (\bibinfo {year} {2012})}\BibitemShut {NoStop}%
\bibitem [{\citenamefont {Baboux}\ \emph {et~al.}(2018)\citenamefont {Baboux},
  \citenamefont {Bernardis}, \citenamefont {Goblot}, \citenamefont {Gladilin},
  \citenamefont {Gomez}, \citenamefont {Galopin}, \citenamefont {Gratiet},
  \citenamefont {Lema\^{i}tre}, \citenamefont {Sagnes}, \citenamefont
  {Carusotto}, \citenamefont {Wouters}, \citenamefont {Amo},\ and\
  \citenamefont {Bloch}}]{Baboux2018}%
  \BibitemOpen
  \bibfield  {author} {\bibinfo {author} {\bibfnamefont {F.}~\bibnamefont
  {Baboux}}, \bibinfo {author} {\bibfnamefont {D.~D.}\ \bibnamefont
  {Bernardis}}, \bibinfo {author} {\bibfnamefont {V.}~\bibnamefont {Goblot}},
  \bibinfo {author} {\bibfnamefont {V.~N.}\ \bibnamefont {Gladilin}}, \bibinfo
  {author} {\bibfnamefont {C.}~\bibnamefont {Gomez}}, \bibinfo {author}
  {\bibfnamefont {E.}~\bibnamefont {Galopin}}, \bibinfo {author} {\bibfnamefont
  {L.~L.}\ \bibnamefont {Gratiet}}, \bibinfo {author} {\bibfnamefont
  {A.}~\bibnamefont {Lema\^{i}tre}}, \bibinfo {author} {\bibfnamefont
  {I.}~\bibnamefont {Sagnes}}, \bibinfo {author} {\bibfnamefont
  {I.}~\bibnamefont {Carusotto}}, \bibinfo {author} {\bibfnamefont
  {M.}~\bibnamefont {Wouters}}, \bibinfo {author} {\bibfnamefont
  {A.}~\bibnamefont {Amo}},\ and\ \bibinfo {author} {\bibfnamefont
  {J.}~\bibnamefont {Bloch}},\ }\bibfield  {title} {\bibinfo {title} {Unstable
  and stable regimes of polariton condensation},\ }\href
  {https://doi.org/10.1364/OPTICA.5.001163} {\bibfield  {journal} {\bibinfo
  {journal} {Optica}\ }\textbf {\bibinfo {volume} {5}},\ \bibinfo {pages}
  {1163} (\bibinfo {year} {2018})}\BibitemShut {NoStop}%
\bibitem [{\citenamefont {Vladimirova}\ \emph {et~al.}(2010)\citenamefont
  {Vladimirova}, \citenamefont {Cronenberger}, \citenamefont {Scalbert},
  \citenamefont {Kavokin}, \citenamefont {Miard}, \citenamefont
  {Lema\^{\i}tre}, \citenamefont {Bloch}, \citenamefont {Solnyshkov},
  \citenamefont {Malpuech},\ and\ \citenamefont {Kavokin}}]{Vladimirova2010}%
  \BibitemOpen
  \bibfield  {author} {\bibinfo {author} {\bibfnamefont {M.}~\bibnamefont
  {Vladimirova}}, \bibinfo {author} {\bibfnamefont {S.}~\bibnamefont
  {Cronenberger}}, \bibinfo {author} {\bibfnamefont {D.}~\bibnamefont
  {Scalbert}}, \bibinfo {author} {\bibfnamefont {K.~V.}\ \bibnamefont
  {Kavokin}}, \bibinfo {author} {\bibfnamefont {A.}~\bibnamefont {Miard}},
  \bibinfo {author} {\bibfnamefont {A.}~\bibnamefont {Lema\^{\i}tre}}, \bibinfo
  {author} {\bibfnamefont {J.}~\bibnamefont {Bloch}}, \bibinfo {author}
  {\bibfnamefont {D.}~\bibnamefont {Solnyshkov}}, \bibinfo {author}
  {\bibfnamefont {G.}~\bibnamefont {Malpuech}},\ and\ \bibinfo {author}
  {\bibfnamefont {A.~V.}\ \bibnamefont {Kavokin}},\ }\bibfield  {title}
  {\bibinfo {title} {Polariton-polariton interaction constants in
  microcavities},\ }\href {https://doi.org/10.1103/PhysRevB.82.075301}
  {\bibfield  {journal} {\bibinfo  {journal} {Phys. Rev. B}\ }\textbf {\bibinfo
  {volume} {82}},\ \bibinfo {pages} {075301} (\bibinfo {year}
  {2010})}\BibitemShut {NoStop}%
\bibitem [{\citenamefont {Takemura}\ \emph {et~al.}(2014)\citenamefont
  {Takemura}, \citenamefont {Trebaol}, \citenamefont {Wouters}, \citenamefont
  {Portella-Oberli},\ and\ \citenamefont {Deveaud}}]{takemura2014polaritonic}%
  \BibitemOpen
  \bibfield  {author} {\bibinfo {author} {\bibfnamefont {N.}~\bibnamefont
  {Takemura}}, \bibinfo {author} {\bibfnamefont {S.}~\bibnamefont {Trebaol}},
  \bibinfo {author} {\bibfnamefont {M.}~\bibnamefont {Wouters}}, \bibinfo
  {author} {\bibfnamefont {M.~T.}\ \bibnamefont {Portella-Oberli}},\ and\
  \bibinfo {author} {\bibfnamefont {B.}~\bibnamefont {Deveaud}},\ }\bibfield
  {title} {\bibinfo {title} {Polaritonic feshbach resonance},\ }\href@noop {}
  {\bibfield  {journal} {\bibinfo  {journal} {Nature Physics}\ }\textbf
  {\bibinfo {volume} {10}},\ \bibinfo {pages} {500} (\bibinfo {year}
  {2014})}\BibitemShut {NoStop}%
\bibitem [{sup()}]{suppl}%
  \BibitemOpen
  \href@noop {} {}\bibinfo {note} {See Supplemental Material at [URL will be
  inserted by publisher].}\BibitemShut {Stop}%
\bibitem [{\citenamefont {Solnyshkov}\ \emph {et~al.}(2016)\citenamefont
  {Solnyshkov}, \citenamefont {Nalitov},\ and\ \citenamefont
  {Malpuech}}]{Solnyshkov2016}%
  \BibitemOpen
  \bibfield  {author} {\bibinfo {author} {\bibfnamefont {D.~D.}\ \bibnamefont
  {Solnyshkov}}, \bibinfo {author} {\bibfnamefont {A.~V.}\ \bibnamefont
  {Nalitov}},\ and\ \bibinfo {author} {\bibfnamefont {G.}~\bibnamefont
  {Malpuech}},\ }\bibfield  {title} {\bibinfo {title} {{Kibble-Zurek Mechanism
  in Topologically Nontrivial Zigzag Chains of Polariton Micropillars}},\
  }\href {https://doi.org/10.1103/PhysRevLett.116.046402} {\bibfield  {journal}
  {\bibinfo  {journal} {Physical Review Letters}\ }\textbf {\bibinfo {volume}
  {116}},\ \bibinfo {pages} {046402} (\bibinfo {year} {2016})}\BibitemShut
  {NoStop}%
\bibitem [{\citenamefont {Hasan}\ and\ \citenamefont {Kane}(2010)}]{Hasan2010}%
  \BibitemOpen
  \bibfield  {author} {\bibinfo {author} {\bibfnamefont {M.~Z.}\ \bibnamefont
  {Hasan}}\ and\ \bibinfo {author} {\bibfnamefont {C.~L.}\ \bibnamefont
  {Kane}},\ }\bibfield  {title} {\bibinfo {title} {\textit{Colloquium} :
  Topological insulators},\ }\href {https://doi.org/10.1103/RevModPhys.82.3045}
  {\bibfield  {journal} {\bibinfo  {journal} {Rev. Mod. Phys.}\ }\textbf
  {\bibinfo {volume} {82}},\ \bibinfo {pages} {3045} (\bibinfo {year}
  {2010})}\BibitemShut {NoStop}%
\bibitem [{\citenamefont {Leblanc}\ \emph {et~al.}(2021)\citenamefont
  {Leblanc}, \citenamefont {Malpuech},\ and\ \citenamefont
  {Solnyshkov}}]{Leblanc2021}%
  \BibitemOpen
  \bibfield  {author} {\bibinfo {author} {\bibfnamefont {C.}~\bibnamefont
  {Leblanc}}, \bibinfo {author} {\bibfnamefont {G.}~\bibnamefont {Malpuech}},\
  and\ \bibinfo {author} {\bibfnamefont {D.~D.}\ \bibnamefont {Solnyshkov}},\
  }\bibfield  {title} {\bibinfo {title} {Universal semiclassical equations
  based on the quantum metric for a two-band system},\ }\href
  {https://doi.org/10.1103/PhysRevB.104.134312} {\bibfield  {journal} {\bibinfo
   {journal} {Phys. Rev. B}\ }\textbf {\bibinfo {volume} {104}},\ \bibinfo
  {pages} {134312} (\bibinfo {year} {2021})}\BibitemShut {NoStop}%
\bibitem [{\citenamefont {Tanese}\ \emph {et~al.}(2013)\citenamefont {Tanese},
  \citenamefont {Flayac}, \citenamefont {Solnyshkov}, \citenamefont {Amo},
  \citenamefont {Lemaitre}, \citenamefont {Galopin}, \citenamefont {Braive},
  \citenamefont {Senellart}, \citenamefont {Sagnes}, \citenamefont {Malpuech}
  \emph {et~al.}}]{tanese2013polariton}%
  \BibitemOpen
  \bibfield  {author} {\bibinfo {author} {\bibfnamefont {D.}~\bibnamefont
  {Tanese}}, \bibinfo {author} {\bibfnamefont {H.}~\bibnamefont {Flayac}},
  \bibinfo {author} {\bibfnamefont {D.}~\bibnamefont {Solnyshkov}}, \bibinfo
  {author} {\bibfnamefont {A.}~\bibnamefont {Amo}}, \bibinfo {author}
  {\bibfnamefont {A.}~\bibnamefont {Lemaitre}}, \bibinfo {author}
  {\bibfnamefont {E.}~\bibnamefont {Galopin}}, \bibinfo {author} {\bibfnamefont
  {R.}~\bibnamefont {Braive}}, \bibinfo {author} {\bibfnamefont
  {P.}~\bibnamefont {Senellart}}, \bibinfo {author} {\bibfnamefont
  {I.}~\bibnamefont {Sagnes}}, \bibinfo {author} {\bibfnamefont
  {G.}~\bibnamefont {Malpuech}}, \emph {et~al.},\ }\bibfield  {title} {\bibinfo
  {title} {Polariton condensation in solitonic gap states in a one-dimensional
  periodic potential},\ }\href@noop {} {\bibfield  {journal} {\bibinfo
  {journal} {Nature communications}\ }\textbf {\bibinfo {volume} {4}},\
  \bibinfo {pages} {1749} (\bibinfo {year} {2013})}\BibitemShut {NoStop}%
\bibitem [{\citenamefont {Abbarchi}\ \emph {et~al.}(2013)\citenamefont
  {Abbarchi}, \citenamefont {Amo}, \citenamefont {Sala}, \citenamefont
  {Solnyshkov}, \citenamefont {Flayac}, \citenamefont {Ferrier}, \citenamefont
  {Sagnes}, \citenamefont {Galopin}, \citenamefont {Lema{\^\i}tre},
  \citenamefont {Malpuech} \emph {et~al.}}]{abbarchi2013macroscopic}%
  \BibitemOpen
  \bibfield  {author} {\bibinfo {author} {\bibfnamefont {M.}~\bibnamefont
  {Abbarchi}}, \bibinfo {author} {\bibfnamefont {A.}~\bibnamefont {Amo}},
  \bibinfo {author} {\bibfnamefont {V.}~\bibnamefont {Sala}}, \bibinfo {author}
  {\bibfnamefont {D.}~\bibnamefont {Solnyshkov}}, \bibinfo {author}
  {\bibfnamefont {H.}~\bibnamefont {Flayac}}, \bibinfo {author} {\bibfnamefont
  {L.}~\bibnamefont {Ferrier}}, \bibinfo {author} {\bibfnamefont
  {I.}~\bibnamefont {Sagnes}}, \bibinfo {author} {\bibfnamefont
  {E.}~\bibnamefont {Galopin}}, \bibinfo {author} {\bibfnamefont
  {A.}~\bibnamefont {Lema{\^\i}tre}}, \bibinfo {author} {\bibfnamefont
  {G.}~\bibnamefont {Malpuech}}, \emph {et~al.},\ }\bibfield  {title} {\bibinfo
  {title} {Macroscopic quantum self-trapping and josephson oscillations of
  exciton polaritons},\ }\href@noop {} {\bibfield  {journal} {\bibinfo
  {journal} {Nature Physics}\ }\textbf {\bibinfo {volume} {9}},\ \bibinfo
  {pages} {275} (\bibinfo {year} {2013})}\BibitemShut {NoStop}%
\end{thebibliography}%

\renewcommand{\thefigure}{S\arabic{figure}}
\setcounter{figure}{0}
\renewcommand{\theequation}{S\arabic{equation}}
\setcounter{equation}{0}

\section{Supplemental Materials}

In this Supplemental Material, we present additional results concerning the case of effectively attractive interactions. We also provide the details of the solution of the non-linear Dirac equation.

\subsection{Effectively attractive interactions}

In the main text, we have considered the case of effectively repulsive interactions ($g<0$, $m<0$). Here, we consider the case of effectively attractive interactions arising from the more usual polariton-polariton repulsion $g>0$ because of the  negative effective mass $m<0$. It is known that effectively attractive interactions for a condensate of massive particles in 2D do not lead to a stationary solution in the continuous limit: the condensate either spreads to infinity or collapses to a single point, instead of forming a soliton. This does not prevent the observation of bright solitons in lattices \cite{Lumer2013}, it only makes impossible to describe them in the continuous limit (with an effective mass).

\begin{figure}[tbp]
    \centering
    \includegraphics[width=0.99\linewidth]{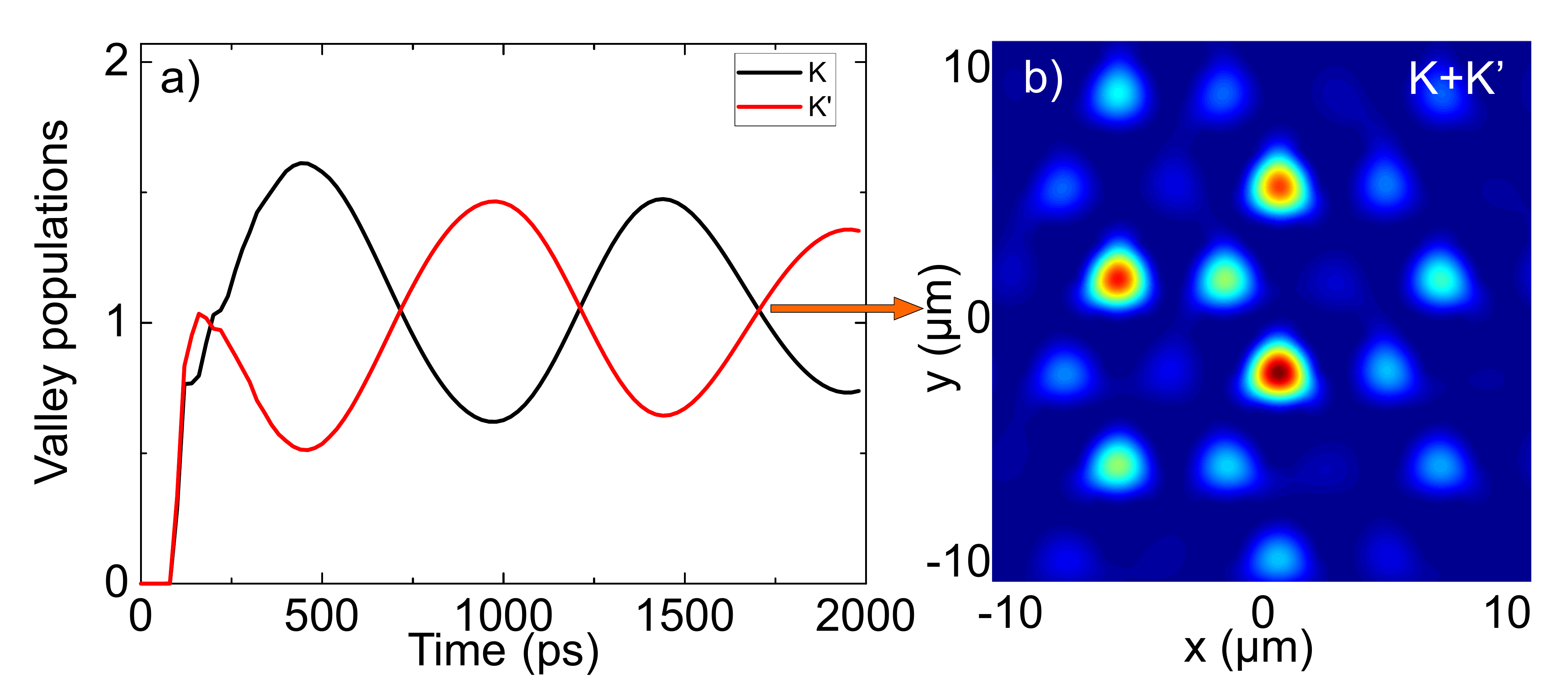}
    \caption{(a) Time dependence of the valley populations exhibiting oscillations induced by the reservoir potential. (b) Spatial image of the condensate in a superposition of valleys, exhibiting a double periodicity.}
    \label{figS1}
\end{figure}

To avoid the problem of condensate spreading or collapse, we consider a finite-size ($40$~$\mu$m) Gaussian reservoir, providing  repulsive interactions $U_R\sim 1$ meV at its center, which are much larger than polariton-polariton interactions $g\left|\psi\right|^2$ and possible disorder effects. Finite spot size favors condensation in negative mass states by confining them under the spot \cite{Jacqmin2014,tanese2013polariton}. Indeed, a repulsive potential is in fact effectively attractive for negative mass particles. So condensation occurs in the K and K' points, at the top of the valence band, in the states showing both the longer lifetime and the best spatial overlap with the reservoir acting on them as an attractive potential. 

Indeed, we observe the formation of a condensate projecting coherently on the six energy degenerate Dirac points with oscillations between the valleys $K$ and $K'$. Contrary to the configuration described in the main text, the relatively small size of the reservoir does not allow to observe multiple valley-polarized domains for this value of reservoir gain. The oscillations between valleys actually occur within the single spatial region.
We have plotted the populations of the two valleys as a function of time in Fig.~\ref{figS1}(a). At the moment of the formation, the populations of the two valleys are approximately equal (in this particular numerical experiment), and then they start to oscillate with a period of about 1~ns. These oscillations are due to the coupling of the two valleys induced by the breaking of the lattice symmetry by the pump, as already observed experimentally at the $\Gamma^\ast$ point in \cite{Milic2018}. Even the slightest displacement or asymmetry of the pump couples the two valleys, and the splitting of the new eigenstates of the trap created by the pump determines the period of the oscillations. These coupled eigenstates are characterized by a pronounced maximum at one of the sites closest to the maximum of the pump (see below). The  oscillations between the valleys with a period of the order of $1$~ns are associated with the oscillations of the particle density in real space between the filled and empty $A$-sites (the $B$-sites always remain empty). This is illustrated by  Fig.~\ref{figS1}(b), showing an equal superposition of the two valleys, where the real space intensity distribution presents a honeycomb lattice with a larger period, meaning that one third of the $A$-sites (forming a "triangular" lattice) remain empty. The period of the oscillations depends on the pump position and on its size. These oscillations can be seen as non-linear Josephson oscillations between the coupled valleys: stronger interactions between condensed particles block the tunneling from one valley to the other because their degeneracy is lifted. This is known as the self-trapping mechanism \cite{abbarchi2013macroscopic}. Oscillations are therefore blocked and condensed particles stay in one valley only.

 \begin{figure}[tbp]
    \centering
    \includegraphics[width=0.99\linewidth]{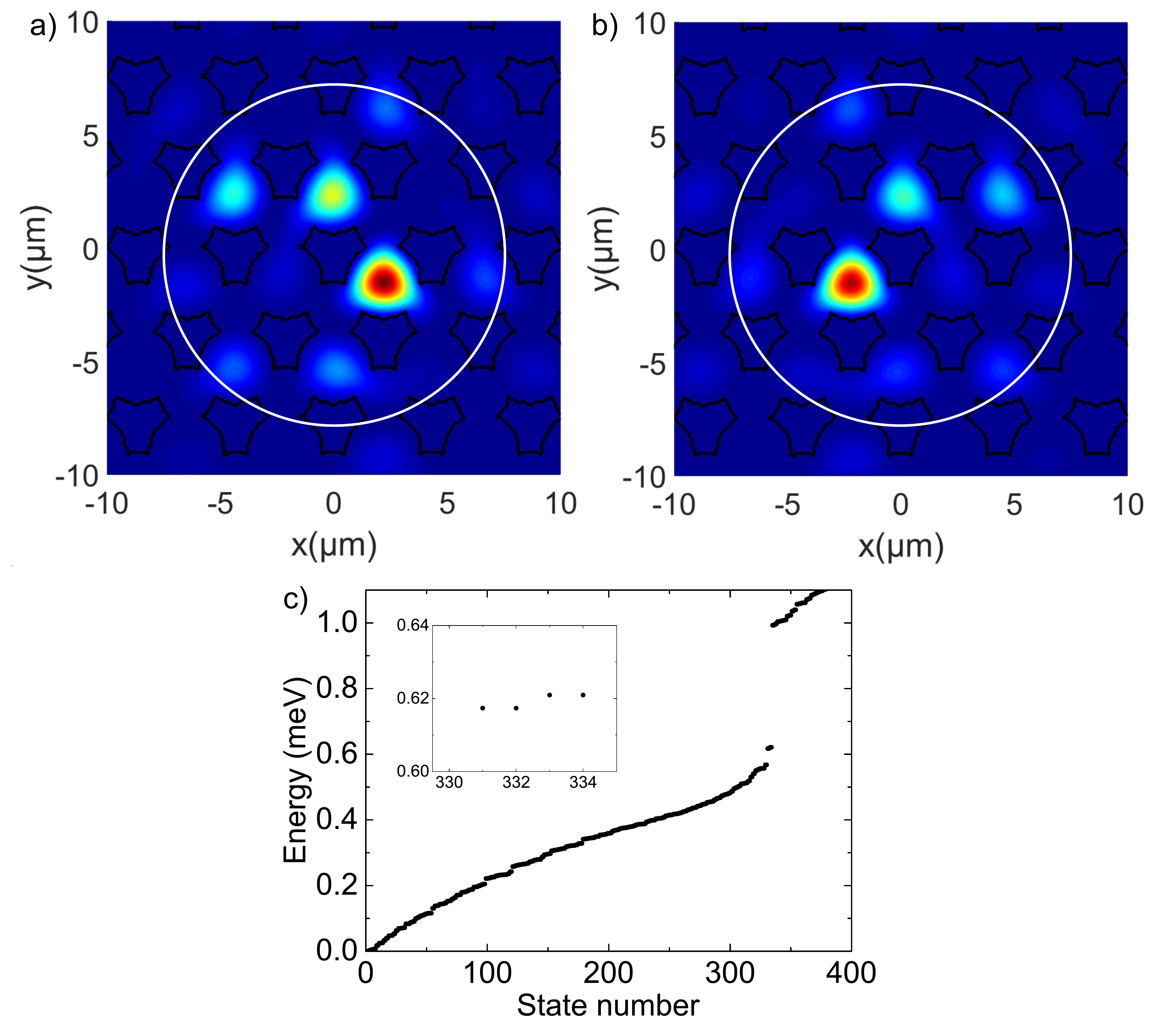}
    \caption{a),b) Particle densities of the two states split by the potential of the pump. The characteristic size of the trap induced by the potential is marked by the white line. c) Energies of the eigenstates of the system, with the two states (a,b) appearing in the gap. The inset shows a zoom on these states, showing the small splitting responsible for valley oscillations. All points are double polarization-degenerate. }
    \label{figS2}
\end{figure}
In order to demonstrate that the oscillations are really due to the splitting of the linear eigenstate of the reservoir potential combined with the lattice, we solve the stationary linear Schr\"odinger equation $\hat{H}\psi_n=E_n\psi_n$, with a potential of 10~$\mu$m size and (for better visibility). This potential is also better centered at one of the lattice sites, so the resulting energy splitting is approximately the same as in Fig.~\ref{figS1}. Interactions, lifetime, and pumping are neglected in this calculation. The results are shown in Fig.~\ref{figS2}: panels (a,b) present the spatial distributions $|\psi(\bm{r})|^2$ of the two almost degenerate states that appear in the gap due to the potential of the pump (reservoir). The energies of some of the eigenstates are shown in panel (c), whose inset presents a zoom on the two localized states appearing in (a,b). From the inset, it is clear that while the states are almost degenerate, this degeneracy is actually removed by the symmetry breaking induced by the slightest shift of the pump with respect to the center of the hexagons. Such symmetry breaking has already been observed experimentally and discussed theoretically in \cite{Milic2018}, at the edge of the upper gap ($\Gamma^\ast$ point). We note that all states are also double polarization-degenerate in this calculation.

The symmetry breaking couples the two valleys. The main visible effect of such coupling is the redistribution of intensity $|\psi(\bm{r})|^2$ in real space. A single valley presents a homogeneous distribution over all sites of the same type $|\psi^K_A|^2=\text{const}$. A superposition of valleys leads to the redistribution of this density, with maxima on some particular sites, closest to the overall minimum of the shifted potential. In particular, the eigenstates shown  in Fig.~\ref{figS2}(a,b) correspond to a superposition $\psi^K+i\psi^{K'}$.

The splitting between the states is controlled by the size of the pump $w$ (it decreases with the increase of $w$) and by its displacement with respect to the center of the hexagon (it increases with the increase of the displacement). The imperfectness of the numerically simulated lattice potential does not allow to reduce this splitting to zero.

\subsection{Solution of the non-linear Dirac equation}
In this section, we demonstrate that the solution (5) given in the main text is indeed the solution of the non-linear Dirac equation to the first order. 
First of all, let's focus on the non-linear term which includes both intra- and inter-valley interaction, but only on the same site:
\begin{widetext}
\begin{equation}
    \left(g\left|\psi_A^K\right|^2+g\left|\psi_A^{K'}\right|^2 \right)\psi_A^K=
    g\left(\left(\frac{1}{2}+ax\right)^2+\left(\frac{1}{2}-ax\right)^2 \right)\left(\frac{1}{2}+ax\right)e^{ik_y y}
    \approx \frac{g}{2}\psi_A^K
\end{equation}
\end{widetext}
A similar result is obtained for the $B$ sites: $\left(g\left|\psi_B^K\right|^2+g\left|\psi_B^{K'}\right|^2 \right)\psi_B^K=\frac{g}{2}\psi_B^K$, because of the absence of a spatial dependence for $\psi_B$ in the first order.
Inserting (4) into (5) gives therefore, for the first line:
\begin{equation}
    \left(\left(\Delta+\frac{g}{2}\right)+\hbar ck_y\right)\left(\frac{1}{2}+ax\right)e^{ik_y y}=E\left(\frac{1}{2}+ax\right)e^{ik_y y},
\end{equation}
where we have used that the term $x k_y$ is of the second order. For the second line we obtain:
\begin{equation}
\left(\hbar c k_y-2\hbar c a-\Delta+\frac{g}{2}\right)\frac{1}{2}e^{ik_y y}=E\frac{1}{2}e^{ik_y y}
\end{equation}
The 3rd and the 1st lines are the same, thanks to the mutual compensation of the changes of sign in the effective field ($-\partial/\partial x$ for $K$ and $+\partial/\partial x$ for $K'$) and of the spatial dependence in the solution (5), containing $+ax$ for $K$ and $-ax$ for $K'$. Similarly, the 4th line is the same as the 2nd one. The system of equations is verified if all 4 lines give the same equation for energy, which allows to determine the parameter $a$ of the solution (5) from the 1st and 2nd line (for example):
\begin{equation}
    \Delta+\frac{g}{2}+\hbar c k_y=-\Delta+2\hbar c a +\frac{g}{2}+\hbar c k_y
\end{equation}
which gives
\begin{equation}
a=\frac{\Delta}{\hbar c}    
\end{equation}
mentioned in the main text.

If we instead insert the solution with an opposite pseudospin:
\begin{equation}
    \left| \psi  \right\rangle \approx  \left( {\begin{array}{*{20}{c}}
{\frac{1}{2} + ax}\\
{-\frac{1}{2} }\\
{\frac{1}{2} - ax}\\
{-\frac{1}{2} }
\end{array}} \right)e^{ik_y y}
\label{badsol}
\end{equation}
it will exhibit an opposite group velocity $v_y=-c$, but this solution requires an opposite sign of $a$:
\begin{equation}
    a=-\frac{\Delta}{\hbar c}
\end{equation}
which means that the order of valleys is inverted with respect to the figure~3 of the main text. Indeed, if we imagine a domain of $K$ embedded in $K'$, a chiral current will flow upward on its right boundary and downwards on its left boundary.

\end{document}